\documentclass[12pt]{article}
\usepackage{amsmath}
\usepackage{graphicx,psfrag,epsf}
\usepackage{enumerate}
\usepackage{subfig}
\usepackage{epstopdf}
\usepackage{natbib}
\usepackage{url} 
\usepackage{epstopdf}
\usepackage{multirow}
\usepackage{color}
\usepackage{mathrsfs}
\usepackage{indentfirst}
\usepackage{algorithm}
\usepackage{algpseudocode}
\usepackage{subfig}
\usepackage{amssymb,amsfonts,amsthm,amsmath}

\DeclareMathOperator*{\argmax}{arg\,max}

\newcommand{\bm}[1]{\mbox{\boldmath$ #1 $\unboldmath}}
\theoremstyle{definition} \newtheorem{prop}{Theorem}

\begin{document}

\def\spacingset#1{\renewcommand{\baselinestretch}%
{#1}\small\normalsize} \spacingset{1}


  \begin{center}
    {\Large\bf Deterministic Sampling of Expensive Posteriors\\ Using Minimum Energy Designs}\\
\vspace{.1in}
V. Roshan Joseph \\
{\small Georgia Institute of Technology, Atlanta, GA 30332} \\
\vspace{.1in}
Dianpeng Wang \\
{\small Chinese Academy of Sciences, Beijing, 100190, China} \\
\vspace{.1in}
Li Gu \\
{\small Wells Fargo \& Company, Charlotte, NC 28202} \\
\vspace{.1in}
Shiji Lv\\
{Peking University, Beijing, 100871, China}\\
\vspace{.1in}
Rui Tuo\\
{\small Chinese Academy of Sciences, Beijing, 100190, China}

\end{center}

\begin{abstract}
Markov chain Monte Carlo (MCMC) methods require a large number of samples to approximate a posterior distribution, which can be costly when the likelihood or prior is expensive to evaluate. The number of samples  can be reduced if we can avoid repeated samples and those that are close to each other. This is the idea behind deterministic sampling methods such as Quasi-Monte Carlo (QMC). However, the existing QMC methods aim at sampling from a uniform hypercube, which can miss the high probability regions of the posterior distribution and thus the approximation can be poor. Minimum energy design (MED) is a recently proposed deterministic sampling method, which makes use of the posterior evaluations to obtain a weighted space-filling design in the region of interest. However, the existing implementation of MED is inefficient because it requires several global optimizations and thus numerous evaluations of the posterior. In this article, we develop an efficient algorithm that can generate MED with few posterior evaluations. We also make several improvements to the MED criterion to make it perform better in high dimensions. The advantages of MED over MCMC and QMC are illustrated using an example of calibrating a friction drilling process.
\end{abstract}

\noindent%

\newpage
\section{Introduction}
\label{2sec:intro}

The main challenge in Bayesian computation is the evaluation of high dimensional integrals arising in Bayesian models. Markov chain Monte Carlo (MCMC) methods are commonly used for this purpose. They work by drawing samples from the posterior and then approximating the integrals using sample averages. Efficient methods for MCMC sampling are proposed in the literature, see \citet{Brooks2011} for a review of different sampling methods. However, these methods can be costly in terms of the number of evaluations made on the posterior distribution. This cost is often neglected, especially when the posterior is easy to evaluate. But when the posterior is complex and expensive to evaluate, the cost becomes appreciable. It is not uncommon for the researchers to wait several hours or even days for the MCMC chain to converge and produce final results. This becomes frustrating for the researcher when he/she has to go back and rerun the chains when minor tweaks are made in the models, which are inevitable at the model building stage.

\begin{figure}
\begin{center}
\begin{tabular}{cc}
\includegraphics[width = 0.45\textwidth]{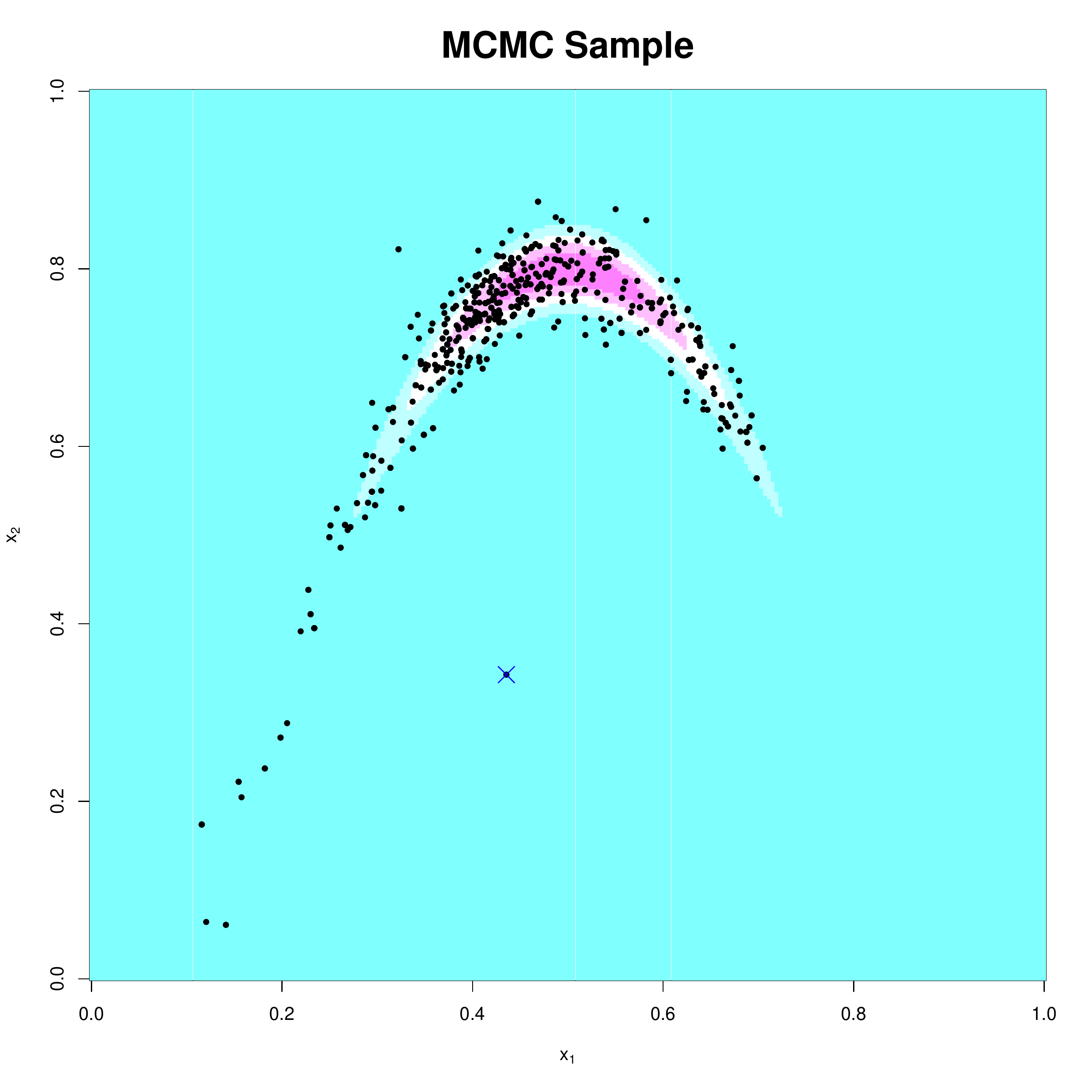} &
\includegraphics[width = 0.45\textwidth]{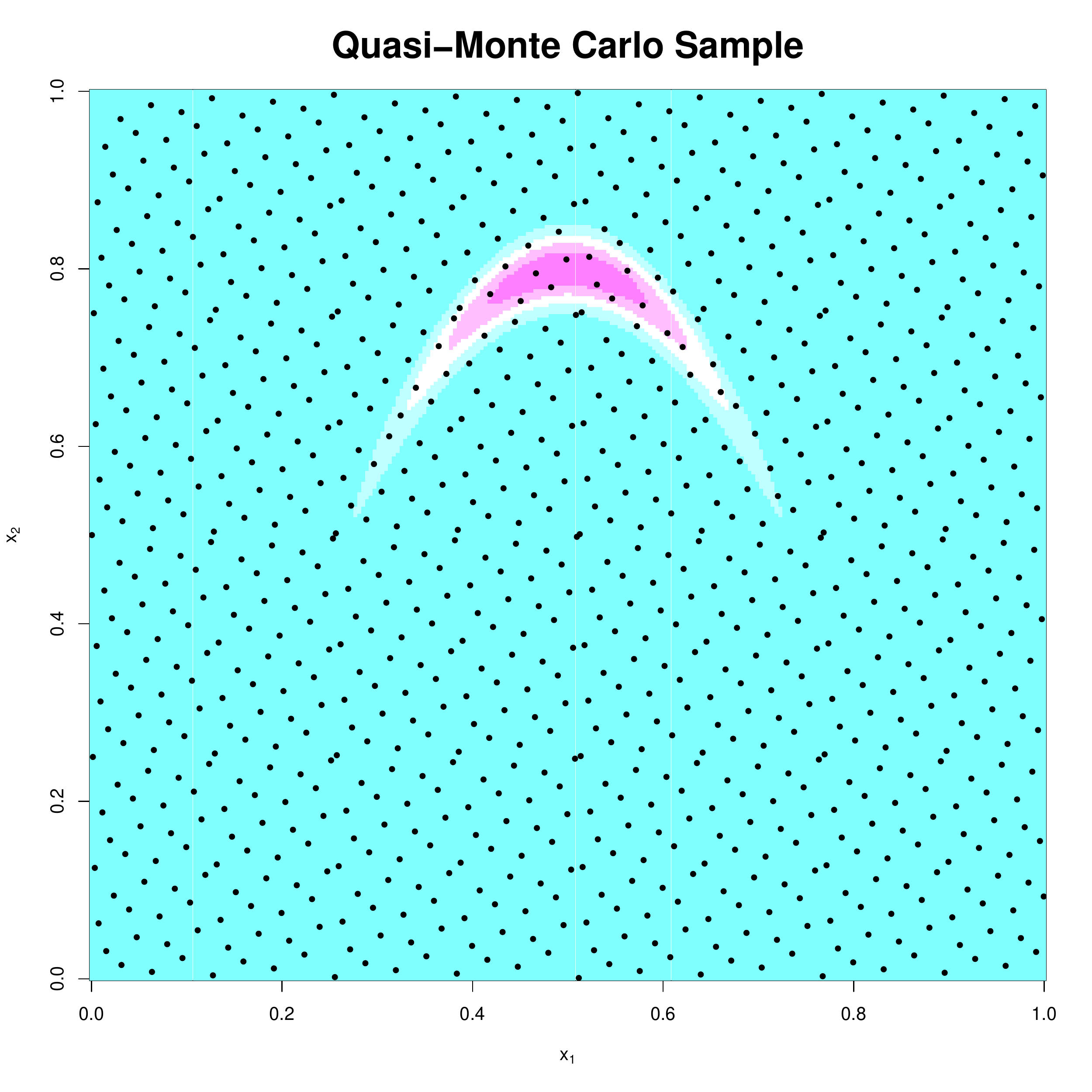}
\end{tabular}
\caption{1000 MCMC samples using robust adaptive Metropolis algorithm  (left) and QMC samples  using Hammersely sequence.}
\label{fig:MCMC-QMC}
\end{center}
\end{figure}

We can overcome the aforementioned problem if we can devise a method that requires only a few evaluations of the posterior. We propose to do this by replacing the ``random'' samples with ``deterministic'' samples. For illustration, consider the banana-shaped density function given by \citep{Haario1999}
\begin{equation}\label{eq:banana}
f(\bm x)\propto \exp\left\{-\frac{1}{2}\frac{x_1^2}{100}-\frac{1}{2} (x_2+.03x_1^2-3)^2\right\}.
\end{equation}
The left panel of Figure \ref{fig:MCMC-QMC} shows 1000 MCMC samples obtained using the robust adaptive sampling proposed by \citet{Vihola2012} and implemented in the R package \emph{adaptMCMC} \citep{adaptMCMC} with a random starting point shown as a ``$\times$'', where the $[0,1]^2$ region corresponds to $[-40,40]\times [-25,10]$ in the original space. We can see that although most of the samples are placed in the high probability region, many samples are repeated or are very close to each other. This is a wastage in terms of function evaluations as they don't give rise to any new information. If we can make the samples as apart as possible and avoid repeated sampling, then we can get more information about the underlying distribution. Quasi-Monte Carlo (QMC) method aims to achieve this through deterministic sampling \citep{Lemieux2009}. The right panel shows 1000 QMC samples obtained using Hammersely sequence. We can see that although the samples are well-spaced, very few fall in the right region where the density is high and thus most of them are wasted. This is a major drawback of QMC methods as they are mainly developed for sampling from unit hypercubes. One recommended strategy in the QMC literature is to transform the samples from the unit hypercube using the inverse of the distribution function. However, this can be done only when the variables are independent with known distribution functions, which is rarely the case in Bayesian problems. Another recommended strategy  is to use importance sampling, but that also has practical limitations because of the challenges in finding a good importance sampling proposal in high dimensions.

The difficulty with the QMC can be avoided if we can  directly sample the points from the posterior distribution and is the main goal of this paper. One such method was proposed by \citet{Joseph2015a} known as Minimum Energy Design (MED). This method draws ideas from experimental designs in computer experiments. Most experimental designs look for uniformity in the region of interest, which is not what we need in this problem. The idea behind MED is to assign some weights in the optimal design criterion so that some areas are preferred over the other areas. \citet{Joseph2015a} showed that by judiciously choosing the weights, the design points can be made to mimic the target distribution. Unfortunately, this idea comes with a price. Choosing the weights and finding the optimal experimental design require tedious global optimizations and numerous evaluations of the posterior distribution making MED noncompetitive to the random sampling-based MCMC methods for most Bayesian problems. This article tries to overcome this serious deficiency of MEDs by proposing an efficient procedure for generating them. Moreover, a generalization of the MED criterion is proposed, which helps improving its performance in high dimensions.

Another recent idea on deterministic sampling is the Stein variational gradient descent proposed by \citet{Liu2016}, but it requires gradient information of the posterior, which can be expensive to compute in high dimensions. It is worth mentioning that the recently developed support points \citep{Mak2017a, Mak2017b} can serve as a much better representative point set than the MED. However, the energy distance criterion used for generating the support points involves the unknown normalizing constant, which doesn't factor out in the optimization as in MED. Thus MED seems to be more useful for generating point sets from computationally expensive probability distributions.

An alternative approach to deal with computationally expensive posteriors is to first approximate the unnormalized posterior with an easy-to-evaluate model and then work on the approximate model instead of the exact posterior. This is the approach taken by many: \citet{Rasmussen2003}, \citet{Bliznyuk2008}, \citet{Fielding2011}, \citet{Bornkamp2011}, and \citet{Joseph2012, Joseph2013}. However, the modeling-based methods are severely limited by the curse of dimensionality. That is, tuning the model becomes extremely difficult in high dimensions leading to poor approximations. Moreover, most modeling methods require an initial set of evaluations in the region of interest. The deterministic samples can be used as an experimental design to build the approximate model and therefore, deterministic sampling techniques can still play an important role even if one wishes to adopt this alternative approach. There has been recent efforts in replacing the global approximation by a series of local approximations within an MCMC sampling scheme \citep{Conrad2016} and adaptively constructing approximate transport maps using MCMC \citep{Parno2016}. We believe that the deterministic sampling technique developed in this paper can make these methods more efficient.

The article is organized as follows.
In Section \ref{2sec:MED}, we review MED and then propose a fast algorithm for constructing MEDs in Section 3.
Some limitations and generalizations of MED are discussed in Section 4. We then illustrate the usefulness of the approach using a computationally expensive posterior example in Section 5. We conclude the article with some remarks in Section 6.

\section{Minimum Energy Designs}
\label{2sec:MED}

Let $\bm D=\{\bm x_1,\ldots, \bm x_n\}$ be the set of deterministic points from the posterior distribution, where each $\bm x_i$ is a $p$-dimensional vector in $\mathbb{R}^p$. It is called a minimum energy design (MED) if it minimizes the total potential energy given by
\begin{equation}\label{eq:energysum}
\sum_{i\ne j} \frac{q(\bm x_i)q(\bm x_j)}{d(\bm x_i,\bm x_j)},
\end{equation}
where $q(\bm x)$ is called a charge function and $d(\bm u,\bm v)$ is the Euclidean distance between the points $\bm u$ and $\bm v$. \citet{Joseph2015a} showed that this is closely related to minimizing
\begin{equation}\label{eq:energy}
E(\bm D)=\max_{i\ne j} \frac{q(\bm x_i)q(\bm x_j)}{d(\bm x_i,\bm x_j)},
\end{equation}
which is more amenable to theoretical analysis. We found that (\ref{eq:energy}) leads to a more computationally stable algorithm and therefore, will use this as the definition of MED from here onwards. Interestingly this design was studied earlier in the literature on sphere packing problems and is known by the name minimal Riesz energy points. \citet{saff2008a, saff2008b} showed that the limiting distribution of MED is given by $q(\bm x)^{-1/(2p)}$.
\citet{Joseph2015a} utilized this property for obtaining a deterministic sample from the probability distribution $f(\bm x)$ by taking $q(\bm x)=1/f^{1/(2p)}(\bm x)$.

Thus, our objective is to find a design that minimizes
\[\max_{i\ne j} \frac{1}{f^{1/(2p)}(\bm x_i)f^{1/(2p)}(\bm x_j)d(\bm x_i,\bm x_j)},\]
or equivalently, a design that maximizes
\begin{equation}\label{eq:MED}
\psi(\bm D)=\min_{i\ne j} f^{1/(2p)}(\bm x_i)f^{1/(2p)}(\bm x_j)d(\bm x_i,\bm x_j).
\end{equation}
An important property that makes this method suitable for Bayesian problems is that we need to know $f(\cdot)$ only up to a constant of proportionality because the  constant does not affect the optimization. So in Bayesian problems, we take $f(\cdot)$ to be the unnormalized posterior. Clearly, an MED will try to place points as apart as possible and in regions where the density is high. Moreover, for finite $n$, the empirical distribution of MED can be considered as an approximation to the target distribution. Thus, MED has all the qualities of a ``deterministic'' sample that we are looking for.

\begin{figure}[h]
\begin{center}
\includegraphics[width = 0.45\textwidth]{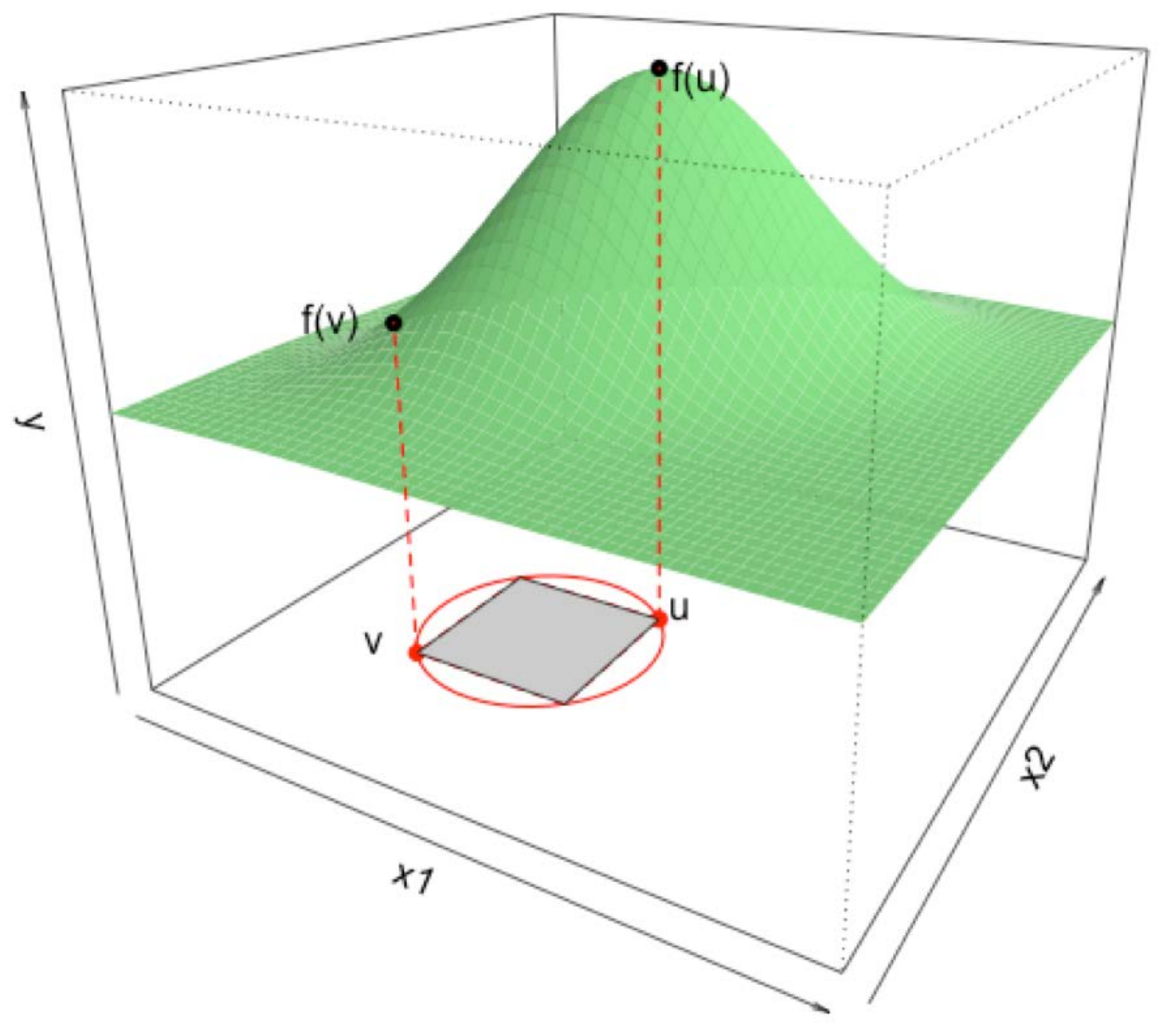}
\caption{Probability-balancing interpretation of MED.}
  \label{fig:prob}
\end{center}
\end{figure}

MED also has a nice probability interpretation. The criterion in (\ref{eq:MED}) is equivalent to
\[\max_{\bm D} \min_{i\ne j} \sqrt{ f(\bm x_i)f(\bm x_j)}V_S(\bm x_i,\bm x_j),\]
where $V_S(\bm x_i,\bm x_j)=\pi^{p/2}/\Gamma (p/2+1) \{d(\bm x_i,\bm x_j)/2\}^p$ is the volume of the sphere with center at $(\bm x_i+\bm x_j)/2$ and passing through the two points $\bm x_i$ and $\bm x_j$. See Figure \ref{fig:prob} for an illustration. The term $\sqrt{ f(\bm x_i)f(\bm x_j)}$ is the geometric mean of the density values at $\bm x_i$ and $\bm x_j$. Thus, $P_{ij}(\bm D)=\sqrt{ f(\bm x_i)f(\bm x_j)}V_S(\bm x_i,\bm x_j)$ is approximately the probability of $\bm x$ falling in the sphere. Let $i^*=arg\min_{j\ne i} P_{ij}(\bm D)$. Then, the MED criterion can be written as
\[\max_{\bm D} \min_{i=1:n} P_{ii^*}(\bm D).\]
Now maximizing the minimum probability will tend to make the probabilities $P_{ii^*}(\bm D)$  equal for all $i=1,\ldots,n$. Thus, roughly speaking, an MED tries to balance the probabilities among adjacent points of the design.

\section{A FAST ALGORITHM FOR GENERATING MED}
Maximizing $\psi(\bm D)$ in \eqref{eq:MED} to find an MED is not an easy problem. \citet{Joseph2015a} proposed a one-point-at-a-time greedy algorithm. The idea is to start with a point $\bm x_1$ and generate $\bm x_2, \bm x_3, \ldots$ sequentially. Thus, the $j$th design point is given by
\begin{equation}\label{eq:GreedyAlgMED}
\bm x_{j} = \argmax _{\bm x} \min_{i=1:(j-1)} f^{1/(2p)}(\bm x)f^{1/(2p)}(\bm x_i)d(\bm x,\bm x_i),
\end{equation}
which is obtained using Generalized Simulated Annealing (GSA) algorithm. Extensive simulations conducted by \citet{Joseph2015a} showed that the algorithm works well as long as $\bm x_1$ is a ``good'' point of the posterior distribution such as posterior mode. However, the mode is difficult to find when the posterior is expensive to evaluate. Moreover, each step of the algorithm requires a global optimization and numerous evaluations of the density $f(\cdot)$, which somewhat defeats the original motivation for developing a deterministic sampling method. For example, generating 100 MED points for the banana-shaped density in (\ref{eq:banana}) using GSA required 43,165 evaluations of the density, which is unacceptably high for a two-dimensional problem. \citet{Joseph2015a} tried to overcome this issue by approximating $f(\cdot)$ using a Gaussian Process (GP) model at each step. However, fitting a global GP model at each step is again a time consuming step making the method non-competitive to both the MCMC and QMC. In this section, we propose an efficient algorithm to generate an MED that overcomes this major limitation.

First assume that, after proper re-scaling, the support of the distribution is a unit hypercube $\mathcal{X}=[0,1]^p$. This re-scaling could be based on the prior distribution or based on the experimenter's prior belief about the limits of each variable. This step looks like the requirement in  QMC, but we will see that because of the sequential nature, the proposed algorithm doesn't suffer as much as the QMC. Now consider an annealed version of the unnormalized posterior density:
\begin{equation}
f^{\gamma}(\bm x),\;\; \gamma \in [0,1],\;\;\bm x\in \mathcal{X}=[0,1]^p,
\end{equation}
where $\gamma=1$ gives the target distribution and $\gamma=0$ gives a uniform distribution in $\mathcal{X}$. Since we can obtain a good point set in $\mathcal{X}$ using the existing QMC methods, we start from $\gamma=0$ and slowly increase it to 1. A major advantage of this approach is that the sampling can explore a multi-modal density in a much better way, which is widely used in optimization and MCMC \citep{Neal1996}.

Suppose we use $K$ steps with $\gamma_k=0,1/(K-1),2/(K-1),\ldots,1$. At each step, we generate $n$ MED points as follows. Let $\bm D_k$ be the MED at the $k$th step and $\bm f_k$ the corresponding  density evaluations. The main idea is to construct $\bm D_{k+1}$ by generating one new point from each of the $n$ points in $\bm D_k$. Thus, at each step, there will be only $n$ new evaluations of the density, which ensures that the total number of evaluations is only $Kn$. A new point at each step is generated as follows. Let $L_{jk}$ denotes a local region around $\bm x_j\in \bm D_k$, $j=1,\ldots,n$. Then the new point is obtained using (\ref{eq:GreedyAlgMED}), which can be equivalently written as
\[
\bm x_j^{new} = \argmax _{\bm x\in L_{jk}} \min_{i=1:(j-1)} \gamma_k \log f(\bm x)+ \gamma_k \log f(\bm x_i)+ 2p \log d(\bm x,\bm x_i).
\]
The use of logarithm ensures numerical stability because the density values can become extremely small. The foregoing step cannot be implemented easily because it requires an optimization albeit in a smaller region. We therefore replace $f(\cdot)$ with an easy-to-evaluate approximation $\hat{f}^{(jk)}(\cdot)$ in $L_{jk}$. Unlike the global approximation needed in \citet{Joseph2015a}, we only need a local approximation of $f(\cdot)$ in $L_{jk}$ and thus, we can use a much cheaper approximation technique. In our implementation, we have used the limit kriging predictor \citep{Joseph2006} using a Gaussian correlation function with a pre-specified value for the correlation parameter. The advantage of limit kriging over the other kriging methods is that it is more robust to the misspecification of the correlation parameters and thus it can be used without much tuning. Thus, in generating
\begin{equation}\label{eq:xjnew}
\bm x_j^{new} = \argmax _{\bm x\in L_{jk}} \min_{i=1:(j-1)} \gamma_k \log \hat{f}^{(jk)}(\bm x) + \gamma_k \log f^{(jk)}(\bm x_i^{new}) + 2p \log d(\bm x,\bm x_i),
\end{equation}
we don't make any new density evaluations. However, the optimization step, even if it is local, can be time consuming. Therefore, we use a space-filling design in $L_{jk}$ and choose the best point in the design according to the criterion in (\ref{eq:xjnew}). This simplification has another major advantage. We can choose the space-filling design in such a way that it is away from the already evaluated points, making each new evaluation useful for learning about the density. We also found it useful to add a few linear combinations of the adjacent points in $\bm D_k$, with weights chosen randomly in $[-.5,1.5]$. Once we obtain $\bm D_{k+1}^{new}=\{\bm x_1^{new}, \ldots,\bm x_n^{new}\}$ and the corresponding evaluations $\{f(\bm x_1^{new}),\ldots,f(\bm x_n^{new})\}$, we obtain $\bm D_{k+1}$ using the greedy algorithm
\begin{equation}
\bm x_{j} = \argmax _{\bm x\in \bm C_{k+1}} \min_{i=1:(j-1)} \gamma_k \log f(\bm x) + \gamma_k \log f(\bm x_i) + 2p \log d(\bm x,\bm x_i),
\end{equation}
for $j=2,\ldots,n$ , where the candidate set
\[\bm C_{k+1}=\bm C_k \cup \bm D_{k+1}^{new}\] with $\bm C_1=\bm D_1$ and $\bm x_1=arg\max_{\bm x\in \bm C_{k+1}}\log f(\bm x)$. Note that there are no new evaluations made for doing this because we search over only the set of points that is already evaluated.

\begin{figure}[h]
\begin{center}
\includegraphics[width = .9\textwidth,height=.69\textwidth]{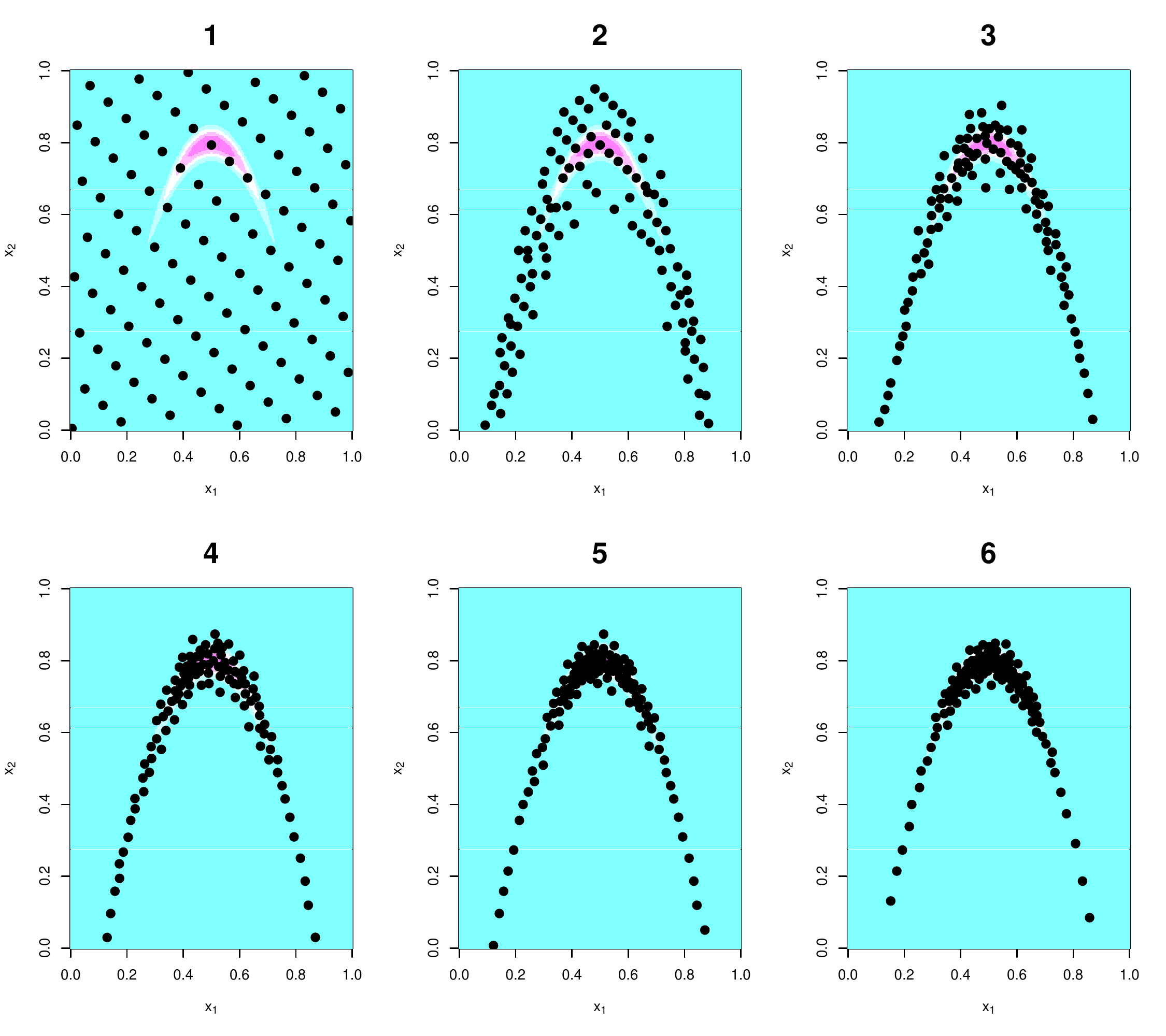}
\caption{Progression of MED points with $\gamma_k=(k-1)/5$ for $k=1,2,\ldots,6$ are shown.}
  \label{fig:banana1-6}
\end{center}
\end{figure}

Consider again the banana-shaped density discussed in the introduction. Let $n=109$ and $K=6$. We use the fast component-by-component construction of lattice rule \citep{Nuyens2006} to obtain $\bm D_1$, which is shown in the first panel of Figure \ref{fig:banana1-6}. The progression of the MED points as $\gamma$ increases to 1 is shown in the same figure. We can see that the final design looks good and captures the density better than the 1000 MCMC and QMC samples in Figure \ref{fig:MCMC-QMC}. Moreover, the total number of density evaluations was only $109\times 6=645$, which is almost two orders of magnitude smaller than the number of evaluations using GSA.

\section{IMPROVEMENTS TO MED}

We observed a few limitations of MED when dealing with uniform distributions and highly correlated distributions. We discuss these limitations in this section and propose some remedies.

\subsection{Uniform Distributions}

\begin{figure}
\begin{center}
\includegraphics[width = .9\textwidth]{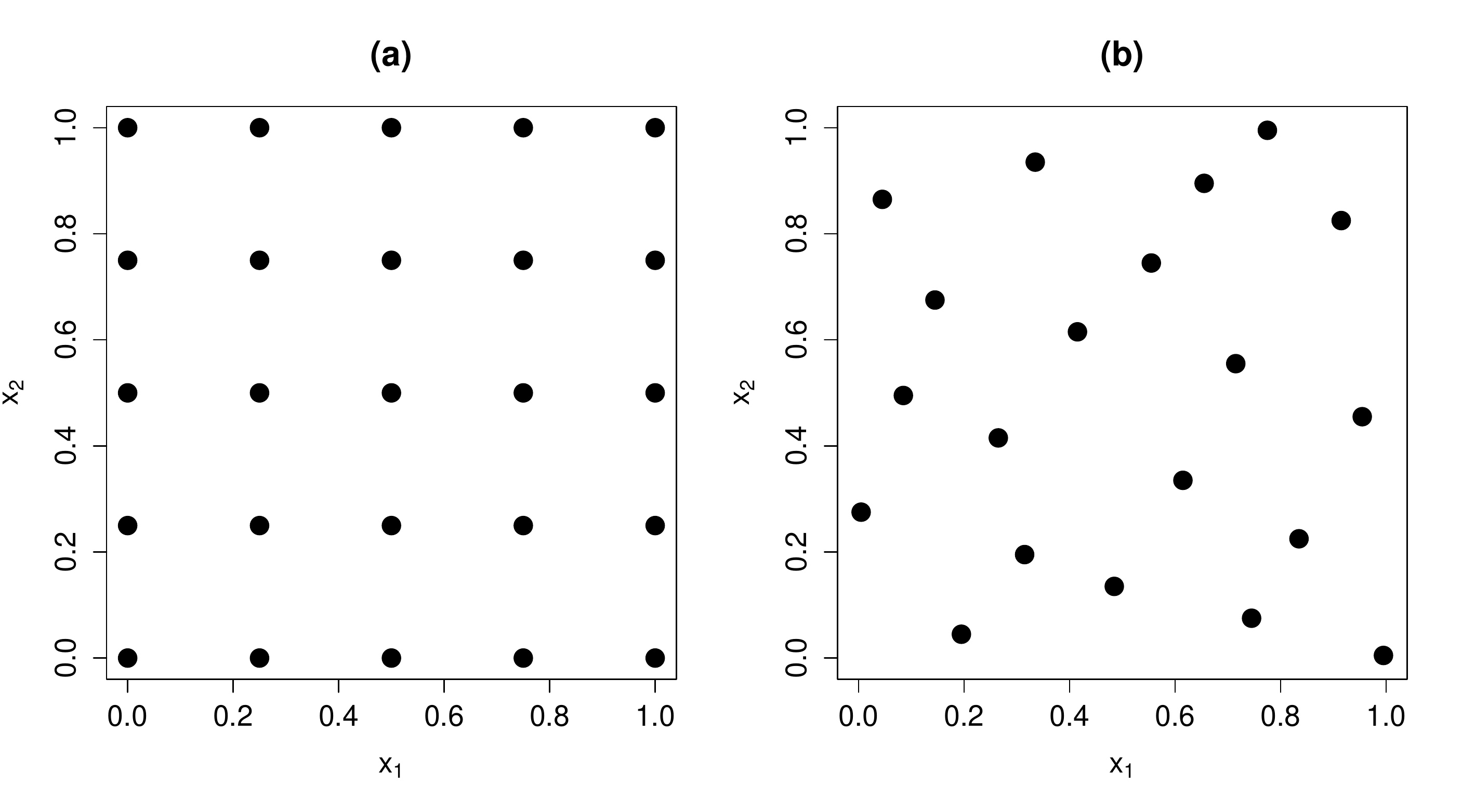}
\caption{25-run MED for uniform distribution with (a) $s=2$ and (b) $s=0$ in (\ref{eq:MEDs}).}
\label{fig:unifrom}
\end{center}
\end{figure}

Figure \ref{fig:unifrom}(a) shows a 25-point MED for a uniform distribution in $[0,1]^2$. This is a full factorial design with five levels for each factor. This structure of the design is expected because the MED reduces to a maximin distance design \citep{Johnson1990} when $f(\cdot)$ is uniform. A factorial-type design is not good in high dimensions because the number of projected points in each dimension from an $n$-run design reduces to no more than $\lceil n^{1/p} \rceil$. Projections can be important because we need to compute marginal distributions and other summary measures of the posterior distribution, which involve high-dimensional integrals. Therefore, when approximating these integrals using sample averages, the error rate will be of the order of $O(n^{-1/p})$, which is worse than the Monte Carlo error rate of $O(n^{-1/2})$ when $p>2$.

\graphicspath{{./Figures/}}

Here is an idea to overcome this problem. Define a generalized distance
\begin{equation}\label{eq:ds}
d_s(\bm u,\bm v)=\left(\frac{1}{p}\sum_{l=1}^p|u_l-v_l|^s\right)^{1/s},
\end{equation}
where $s>0$. For $s<1$, $d_s(\cdot,\cdot)$ is not a metric, but as we show below that it has the desirable properties that are needed to achieve our objectives. Using this generalized distance, the MED criterion becomes
\begin{equation}\label{eq:MEDs}
\max_{\bm D}\psi(\bm D)=\max_{\bm D}\min_{i\ne j} f^{1/(2p)}(\bm x_i)f^{1/(2p)}(\bm x_j)d_s(\bm x_i,\bm x_j).
\end{equation}
Denote this design by $MED_s$. The following result shows that the limiting distribution of $MED_s$ is $f(\bm x)$ irrespective of the value of $s \in (0,\infty)$. The proof is quite long and technical, so it is given in the supplementary file associated with this article.

\begin{prop}\label{Th1}
Suppose the charge function $q(\cdot)$ is Lipschitz continuous, i.e.,
$
|q(\bm u)-q(\bm v)|\leq L d(\bm u,\bm v),
$
for $\bm u,\bm v\in\mathcal{X}=[0,1]^p$ and a constant $L>0$.
Let $D^*=\{\bm x_1^*,\ldots,\bm x_n^*\}$ be an $n$-point minimum energy design using (\ref{eq:MEDs}) with the smallest index and $\mathscr{B}$ be the Borel $\sigma$-algebra of $\mathcal{X}$. Define the following probability measures on $(\mathcal{X},\mathscr{B})$:
\begin{eqnarray}
\mathcal{P}_{n}(A)=\frac{\mathrm{card}\{\bm x_i^*:1\leq i\leq n, \bm x_i^*\in A\}}{n},\text{ for any } A\in \mathscr{B}.
\end{eqnarray}
Then there exists a probability measure $\mathcal{P}$ such that $\mathcal{P}_{n}$ converges to $\mathcal{P}$ weakly for all fixed $s\in(0,\infty)$  as $n\rightarrow\infty$. Moreover, $\mathcal{P}$ has a density proportional to $1/q^{2p}(\bm x)$ over $\mathcal{X}$ .
\end{prop}

When $s\rightarrow 0$, the criterion becomes
\begin{equation}\label{eq:MED0}
\max_{\bm D}\min_{i\ne j} f^{1/(2p)}(\bm x_i)f^{1/(2p)}(\bm x_j)\prod_{l=1}^p|x_{il}-x_{jl}|^{1/p}.
\end{equation}
Now for $f(\bm x)=1$, the criterion is to maximize $\min_{i\ne j}\prod_{l=1}^p|x_{il}-x_{jl}|^{1/p}$. Figure \ref{fig:unifrom}(b) shows the 25-point $MED_0$, which clearly has better projections than the original MED. The product measure ensures that no two points can have the same coordinate. Thus, the design will project onto $n$ different points in each dimension, a property shared by the popular Latin hypercube designs \citep{Santner2003}. In fact, the criterion in (\ref{eq:MED0}) for $f(\bm x)=1$ is a limiting case of the MaxPro design criterion proposed by \citet{Joseph2015b}. The Latin hypercube and MaxPro designs have much better centered $L_2$ discrepancy ($CL_2$) measures \citep{Hickernell1998} than factorial-type designs and thus, are expected to perform much better in high dimensions. We have not yet established the theoretical convergence rate for the integration errors of these new designs, but our investigation with similar designs show a rate better than the MC rate by a $\log n$ factor \citep{Mak2017a, Mak2017b}.

The MED criterion in (\ref{eq:MED0}) can also be given a probabilistic interpretation. It can be written as
\[\max_{\bm D} \min_{i\ne j} \sqrt{ f(\bm x_i)f(\bm x_j)}V_R(\bm x_i,\bm x_j),\]
where $V_R(\bm x_i,\bm x_j)=\prod_{l=1}^p|x_{il}-x_{jl}|$ is the volume of the hyper-rectangle, which has $\bm x_i$ and $\bm x_j$ at the two opposite corners. See Figure \ref{fig:prob} for an illustration. Thus, the same probability-balancing interpretation as in the Euclidean case holds for this criterion as well, which can be obtained by replacing the hyper-sphere volume element with the hyper-rectangle volume element.

\subsection{Correlated Distributions}
\begin{figure}
\begin{center}
\begin{tabular}{cc}
\includegraphics[width = 0.45\textwidth]{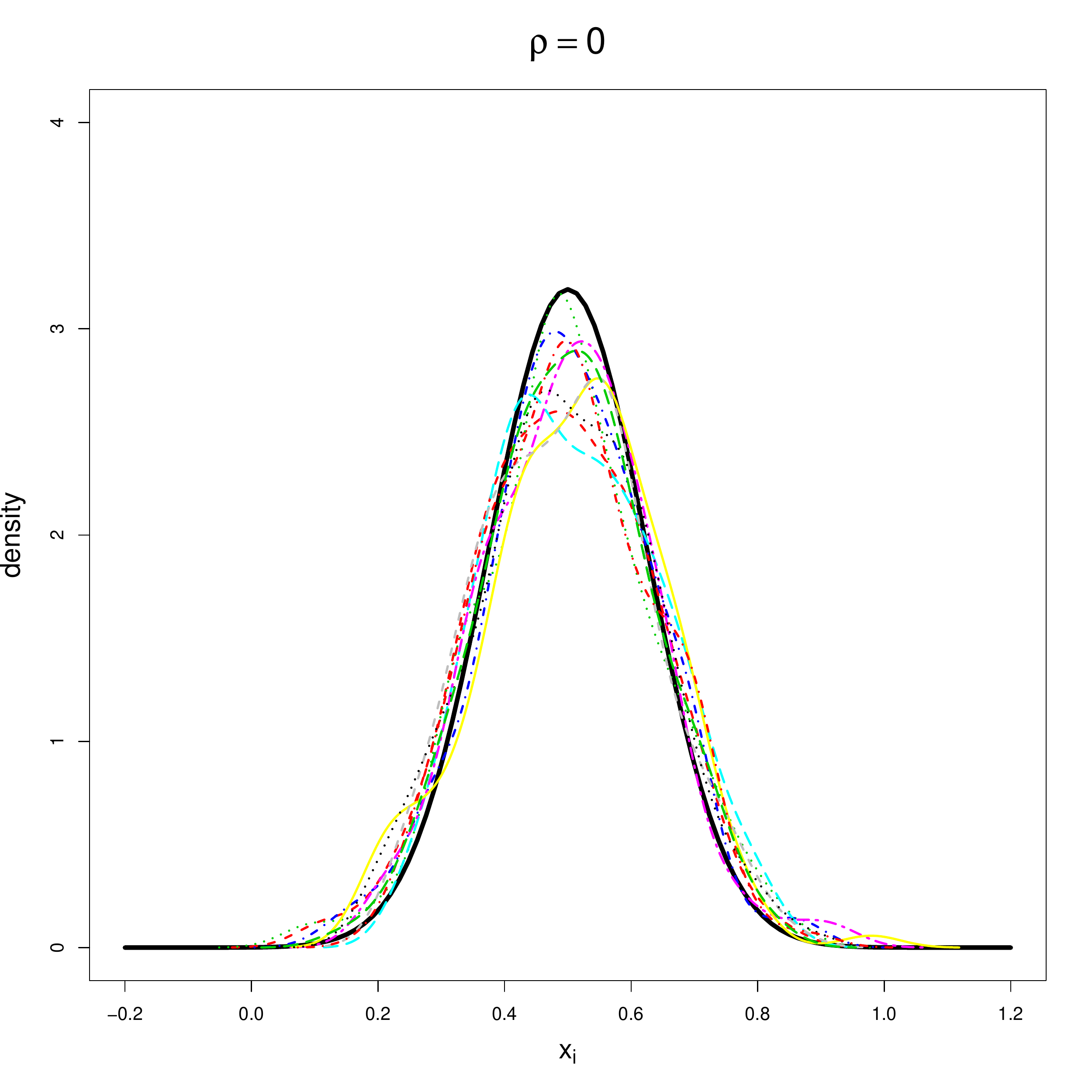} &
\includegraphics[width = 0.45\textwidth]{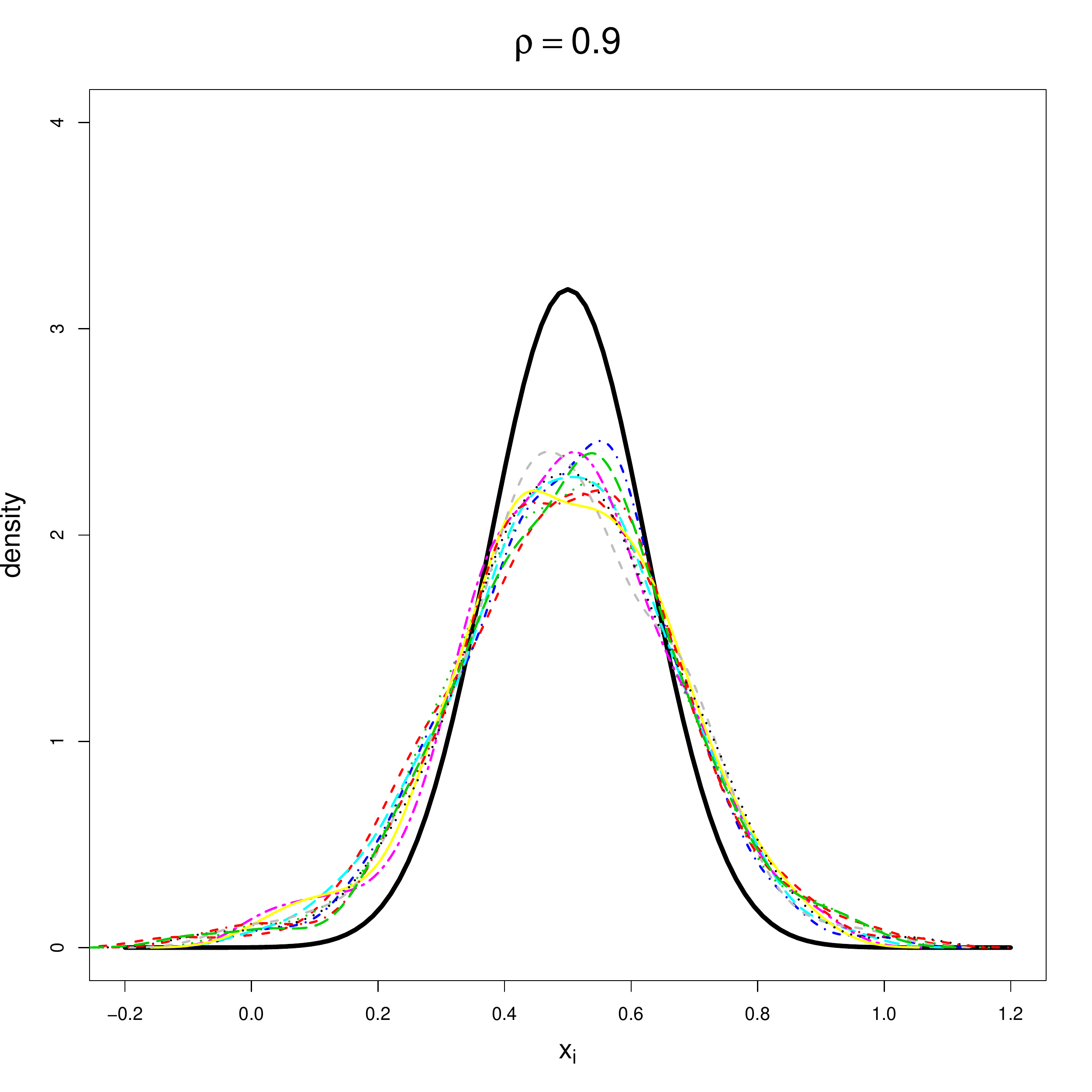}
\end{tabular}
\caption{Estimated marginal distributions from a 149-point MED of a 10-dimensional independent normal distribution (left) and dependent normal distribution with $\rho=0.9$. The true marginal distribution is shown as a thick solid black line.}
\label{fig:normal10}
\end{center}
\end{figure}

Consider a multivariate normal density with mean at $\bm {.5}=(0.5,\ldots,0.5)'$ and a first order auto-regressive correlation structure: $\bm x\sim N(\bm {.5},\sigma^2\bm R)$, where $\sigma=1/8$, $\bm R_{ij}=\rho^{|i-j|}$ for $\rho\in [0,1]$ and $i,j=1,\ldots,p$. The left panel in Figure \ref{fig:normal10} shows the marginal distributions of 149-point MED of a 10-dimensional multivariate normal density with $\rho=0$. We can see that the marginal distributions agree reasonably well with the true marginal density $N(.5,1/8)$, which is shown as a thick solid black line. Now consider the correlated case with $\rho=0.9$. The marginal distributions of 149-point MED is shown in the right panel of Figure \ref{fig:normal10}. We can see that they are more dispersed than the true marginal distribution. We will show below that this problem is not due to the algorithm, but with the MED criterion.

Let $\bm \Sigma=\sigma^2\bm R$ and $f(\bm x)=\exp\{-.5(\bm x-\bm {.5})'\bm \Sigma^{-1}(\bm x-\bm {.5})\}$, omitting the normalizing constant. Consider a two-point MED. By symmetry, they should be of the form $\bm x_1=\bm {.5}-\bm u$ and $\bm x_2=\bm {.5}+\bm u$, where $\bm u$ maximizes
\begin{eqnarray*}
\log\psi(\bm D)&=&.5\log f(\bm {.5}-\bm u)+.5 \log f(\bm {.5}+\bm u)+p\log d(\bm {.5}-\bm u,\bm {.5}+\bm u)\\
&=&-.5\bm u'\bm \Sigma^{-1}\bm u+\frac{p}{2}\log (4\bm u'\bm u).
\end{eqnarray*}
Differentiating and equating to 0, we find that $\bm u$ should satisfy
\[\left(\bm \Sigma-\frac{\bm u'\bm u}{p}\bm I\right)\bm u=0.\]
Now when $\rho=0$, we can see that the solution can be any point in the surface of the sphere $\bm u'\bm u=p\sigma^2$. One particular solution is $\bm u=\pm\sigma \bm 1$. On the other hand, if $\rho>0$, then $\bm u$ is in the eigen direction corresponding to the largest eigen value of $\bm \Sigma$. Consider the extreme case of $\rho=1$. Then, it is easy to show that $\bm u=\pm \sigma\sqrt{p}\bm 1$ is the only solution. Thus, as $p$ increases, MED points move away from the center, which explains the phenomenon observed in Figure \ref{fig:normal10}.

\begin{figure}[h]
\begin{center}
\begin{tabular}{cc}
\includegraphics[width = 0.45\textwidth]{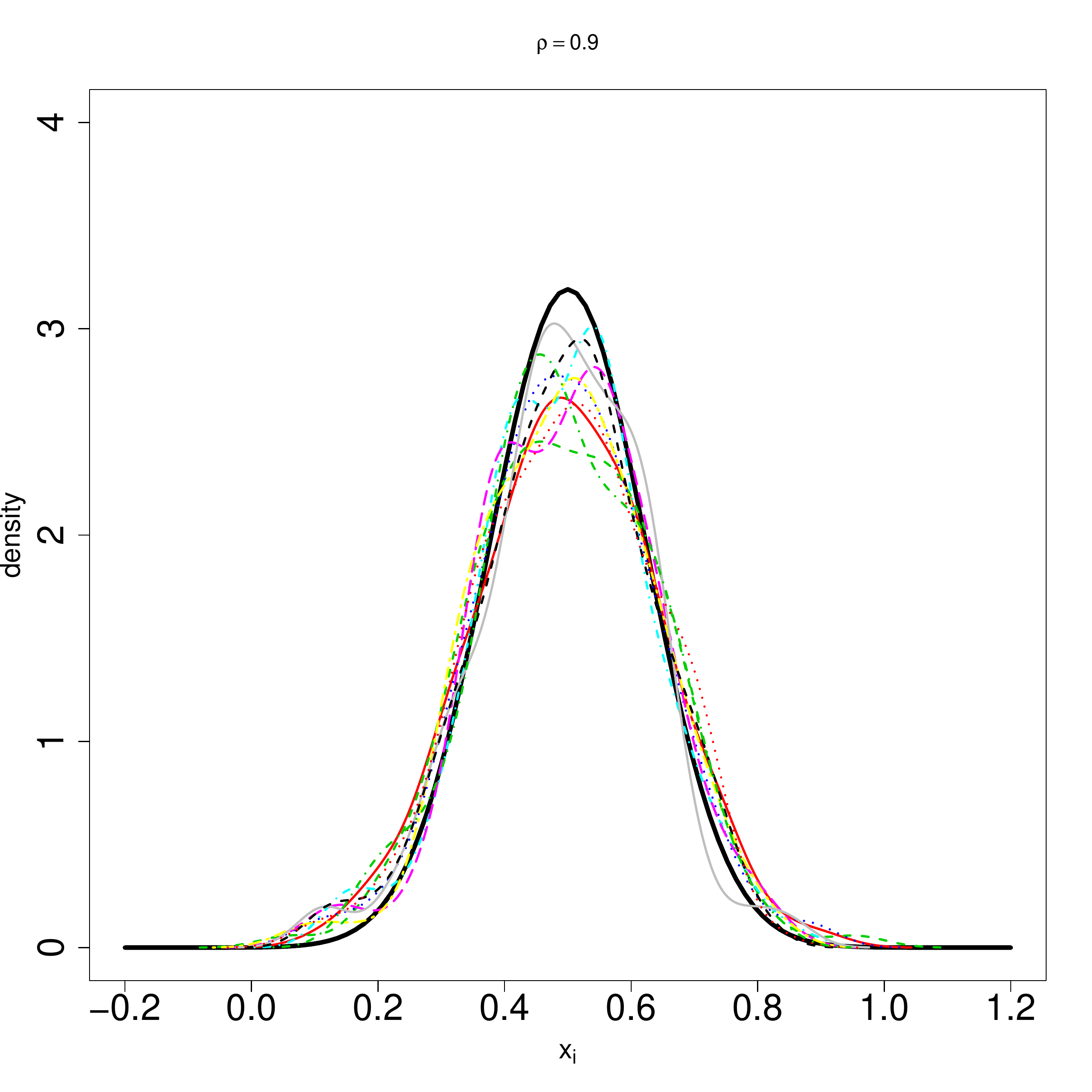} &
\includegraphics[width = 0.45\textwidth]{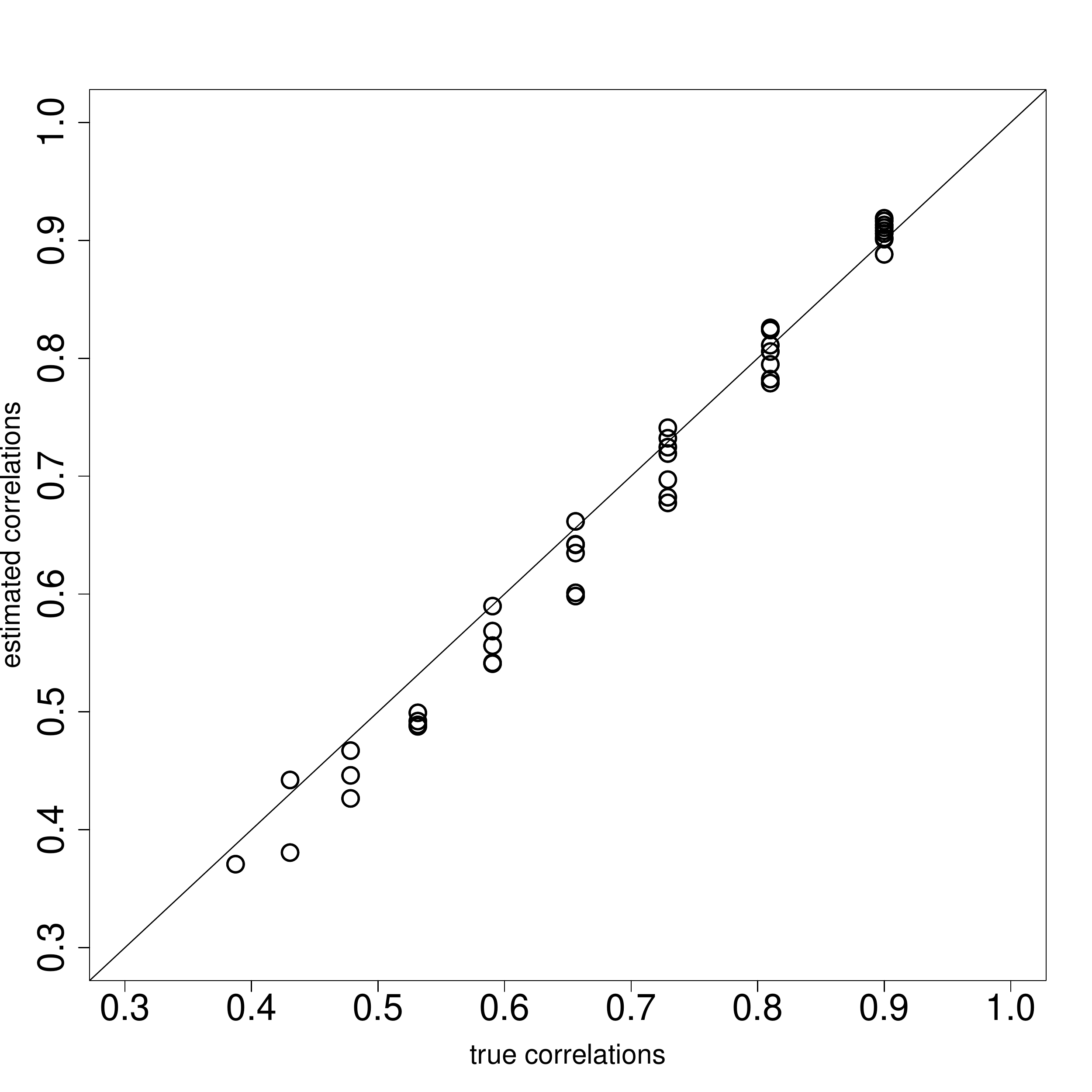}
\end{tabular}
\caption{Estimated marginal distributions from a 149-point MED of a 10-dimensional normal distribution with $\rho=0.9$ using the modified criterion in (\ref{eq:MED_M2}) (left). The estimated correlations are plotted against the true correlations (right).}
\label{fig:normal10_MD}
\end{center}
\end{figure}

Consider a modified MED criterion by replacing the Euclidean distance with the Mahalanobis distance
\begin{equation}\label{eq:MED_M}
\max_{\bm D}\tilde{\psi}(\bm D)=\max_{\bm D}\min_{i\ne j} f^{1/(2p)}(\bm x_i)f^{1/(2p)}(\bm x_j)\tilde{d}(\bm x_i,\bm x_j;\bm \Sigma),
\end{equation}
where $\tilde{d}^2(\bm x_i,\bm x_j;\bm \Sigma)=(\bm x_i-\bm x_j)'\bm \Sigma^{-1}(\bm x_i-\bm x_j)$. The distributional convergence of the MED from (\ref{eq:MED_M}) follows from Theorem 1 using a linear transform of the space. Now for the 2-point MED of a multivariate normal distribution considered earlier, it is easy to show that $\bm u$ can be any point on the ellipsoid $\bm u'\bm \Sigma^{-1} \bm u=p\sigma^2$ irrespective of the value of $\rho$. This makes more sense than the single point solution on the principal eigen direction obtained with the Euclidean distance when $\rho\ne 0$. With this modification, now there is a high chance that the algorithm might choose a solution not too far from the center. However, there is also the danger of choosing the solution too close to the center. Ideally we want to choose a solution that not only maximizes $\tilde{\psi}(\bm D)$, but also $\psi(\bm D)$. So we may consider maximizing
\begin{equation}\label{eq:MED_M2}
w\psi(\bm D)+(1-w)\tilde{\psi}(\bm D),
\end{equation}
for some $w\in (0,1)$. However, this doubles the cost of evaluating the objective function and also introduces the difficulty of choosing a new parameter $w$. Fortunately, there is an easy way to incorporate this into the proposed algorithm without having to go through these additional troubles. Note that the algorithm makes two passes at each step, first to find $\bm D_k^{new}$ and then to find $\bm D_k$. So we simply use $\psi(\bm D)$ in the first step and then use $\tilde{\psi}(\bm D)$ in the second step. This ensures the final design has high objective function values under both criteria. The marginal distributions of the MED points using the modified algorithm is shown in the left panel of Figure \ref{fig:normal10_MD}, which agree reasonably well with the true marginal distribution. The $10\choose 2$ estimated correlations from the MED are plotted against the true correlations in the right panel, which show good agreement. Thus, these two plots together show that the 10-dimensional normal distribution is well-represented by the 149-point MED.

\begin{figure}
\begin{center}
\begin{tabular}{cc}
\includegraphics[width = 0.45\textwidth]{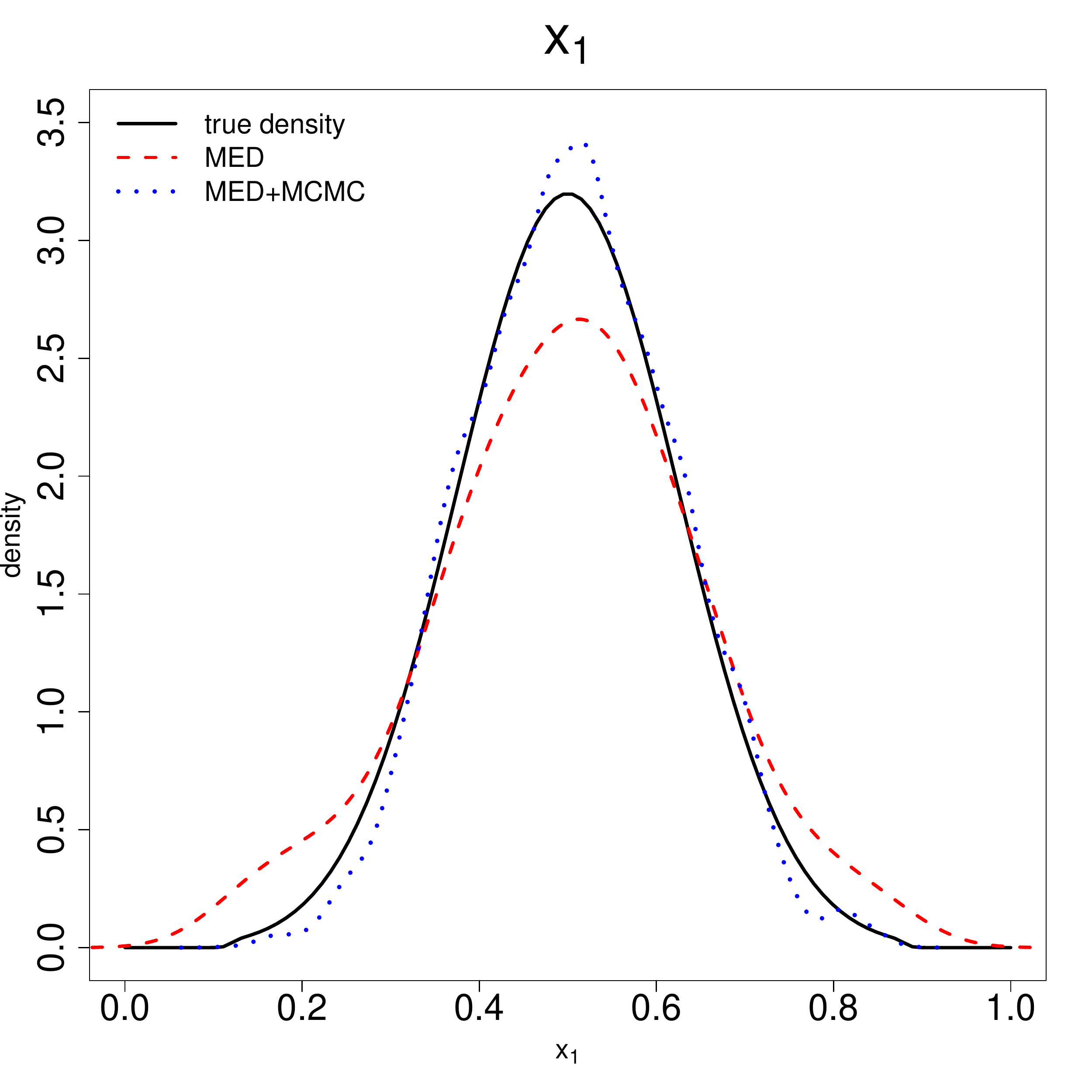} &
\includegraphics[width = 0.45\textwidth]{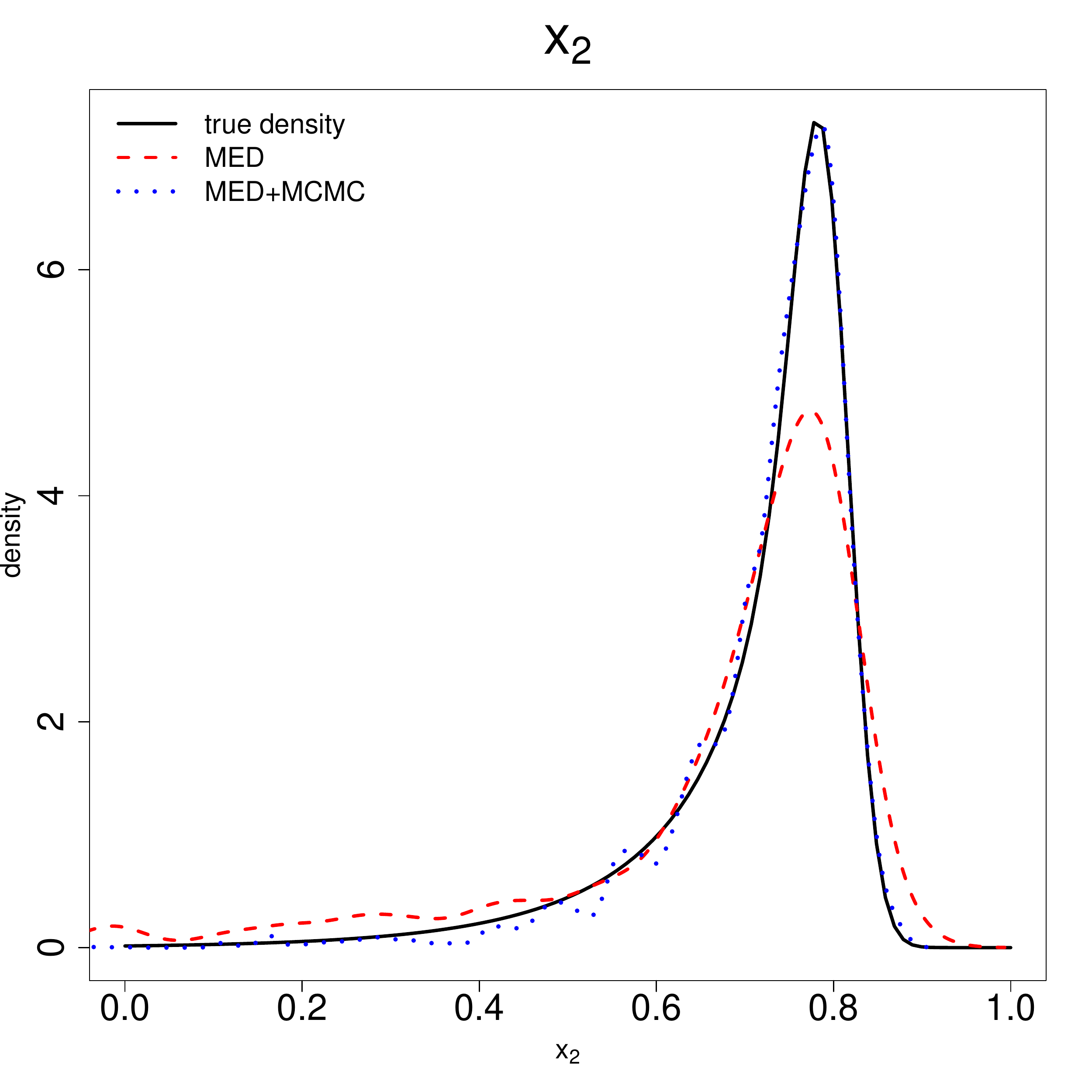}
\end{tabular}
\caption{Estimated marginal distributions from a 109-point MED of the banana-shaped distribution. The true marginal distribution and density from the follow-up MCMC based on the MED points are also shown.}
\label{fig:bananax1x2}
\end{center}
\end{figure}

However, the foregoing improvement on MED using the Mahalanobis distance may not help with more complex distributions such as the banana-shaped density considered earlier. For this distribution, the overall correlation is approximately zero and thus the new criterion reduces to the original criterion. The marginal distributions of the design from the last panel of Figure \ref{fig:banana1-6} is shown in Figure \ref{fig:bananax1x2}. In this case, the true marginal distributions can be obtained through numerical integration and are also shown in the same figure. We can see that the marginal distributions from MED are much more dispersed than the true distributions. One way to fix this issue is to follow-up the MED with an MCMC or a deterministic approximation technique such as DoIt \citep{Joseph2012, Joseph2013}. Consider the option of using MCMC. We first fit a limit kriging predictor on the log-unnormalized posterior using the 654 points. Then we run $n=109$ MCMC chains using Metropolis algorithm starting at each of the $n$ MED points. The size of the $i$th MCMC chain is taken as $\lceil Np_i\rceil$, where $p_i= f(\bm x_i)/\sum_{i=1}^nf(\bm x_i)$ and $N=10,000$. The marginal densities obtained from the resulting MCMC samples are also shown in Figure \ref{fig:bananax1x2}. We can see that they match well with the true distribution. Note that we have not made any new evaluations of the unnormalized posterior for doing this.

\subsection{Improved MED Criterion and Algorithm}
The developments in the previous two subsections suggest an improved version of the MED criterion. We can write (\ref{eq:MED_M}) as
\[\max_{\bm D}\min_{i\ne j} f^{1/(2p)}(\bm x_i)f^{1/(2p)}(\bm x_j)d(\bm \Sigma^{-1/2}\bm x_i,\bm \Sigma^{-1/2}\bm x_j).\]
Now the Euclidean distance $d(\cdot,\cdot)$ can be generalized to obtain
\begin{equation}\label{eq:newMED}
\max_{\bm D}\min_{i\ne j} f^{1/(2p)}(\bm x_i)f^{1/(2p)}(\bm x_j)d_s(\bm \Sigma^{-1/2}\bm x_i,\bm \Sigma^{-1/2}\bm x_j),
\end{equation}
where $d_s(\cdot,\cdot)$ is defined in (\ref{eq:ds}). The only two remaining things to decide are the choices of $\bm \Sigma$ and $s$.

Because we use an iterative algorithm, $\bm \Sigma$ at the $(k+1)$th stage can be estimated using the MED at the $k$th stage. An estimate of $\bm \Sigma$ can be obtained using the sample variance-covariance matrix: $\bm \Sigma_k=\widehat{var}(\bm D_k)$. The inverse of the Hessian of the negative log-likelihood computed at the posterior mode can also be used as an estimate of $\bm \Sigma_k$. This estimate suggests that $\bm \Sigma_{k+1}=\gamma_k/\gamma_{k+1}\bm \Sigma_k$, which makes sense because the variance decreases as $\gamma$ increases. Thus, we use the estimate
\[\bm \Sigma_{k+1}=\frac{\gamma_k}{\gamma_{k+1}}\widehat{var}(\bm D_k).\]

Now consider the choice of $s$. Our experiments with multivariate normal distribution using different values of $s$ gave better results when $s=2$ than $s=0$. We think this  is because it is unlikely that the MED points of a non-uniform distribution will lie parallel to the coordinate axes. Thus, projections seem to be less of a concern with non-uniform distributions. Bayesian asymptotics suggest that the posterior distributions tend to be normal as the sample size increases, so the posterior distribution is likely to be non-uniform. However, when the data is uninformative and when the prior distributions are flat, $s=0$ can perform better. Here is an adaptive choice of $s$ at the $k$th stage:
\[s_k=2\left\{1-\left(\frac{\bm f_{k,min}}{\bm f_{k,max}}\right)^{\gamma_k}\right\},\]
where $\bm f_{k,min}$ and $\bm f_{k,max}$ denote the minimum and maximum values of $\bm f_k$, respectively. We can see that if $\bm f_{k,min}=\bm f_{k,max}$ (i.e., a uniform distribution) or $\gamma_k=0$, then $s_k=0$, but will tend to 2 if $\bm f_{k,min}<<\bm f_{k,max}$ and $\gamma_k$ is large. We can also use a lower and upper quantile of $\bm f_k$ instead of the minimum and maximum values to detect almost flat distributions.

\begin{figure}
\begin{center}
\begin{tabular}{cc}
\includegraphics[width = 0.45\textwidth]{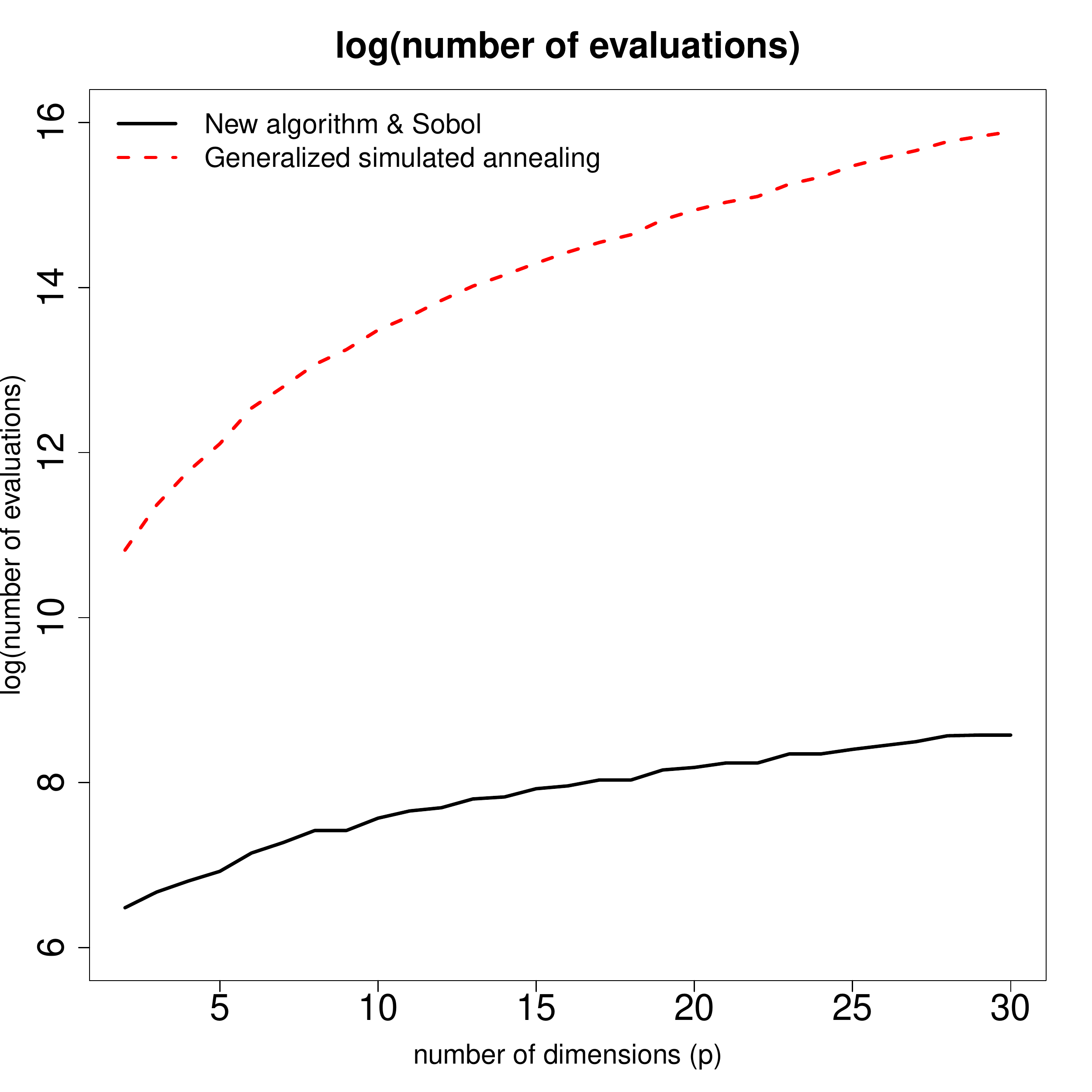} &
\includegraphics[width = 0.45\textwidth]{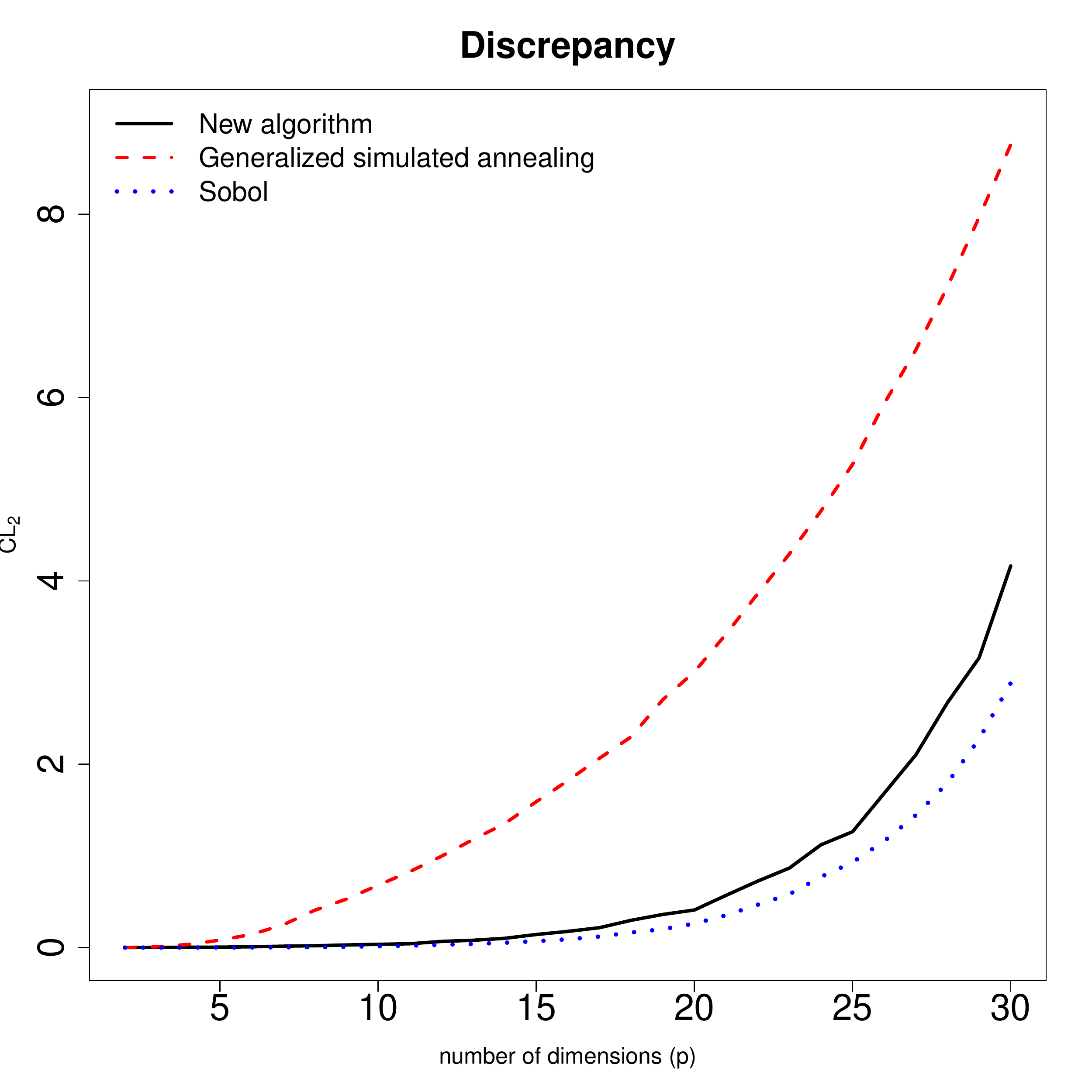}
\end{tabular}
\caption{The logarithm of the number of density evaluations made by the new algorithm and the GSA algorithm are shown in the left panel. The right panel shows the centered $L_2$ discrepancy measure (smaller-the-better).}
\label{fig:sim}
\end{center}
\end{figure}

Consider again the $p$-dimensional normal distribution with $\rho=0.9^{\log p}$ with $p=1,2,...,30$. To use the fast component-by-component construction of the lattice rule, we choose the largest prime number less than $100+5p$ as the size of MED ($n$). There are a few more parameters to choose in the algorithm. We let $K=\lceil 4\sqrt{p}\rceil$ as the number of steps in the algorithm and we choose the closest $n$ points to $\bm x_j$ to define the local region $L_{jk}$. We have used these parameter settings for all the examples presented in this article. The left panel of Figure \ref{fig:sim} shows the logarithm of number of density evaluations made by the new algorithm and the generalized simulated annealing (GSA) algorithm used in \citet{Joseph2015a}. For the 30-dimensional density, the number of evaluations made by the new algorithm is smaller than that of the GSA algorithm by a factor of 1500, which is a substantial saving! In terms of the CPU time, generating $n=241$ MED points for the 30-dimensional density took about 40 minutes using the new algorithm with a total evaluations of 5302. The current implementation is in R and we believe that it can be made an order of magnitude faster by converting it to C++.  The right panel of the Figure \ref{fig:sim} shows the $CL_2$ discrepancy criterion of the MED generated by the two algorithms, which is computed by transforming the point set to a unit hypercube using the normal distribution function. This shows that the MED generated by the new algorithm has much smaller discrepancy and is thus closer to the target distribution. However, the quality of the point set deteriorate as the number of dimensions increases. The $CL_2$ discrepancy of the Sobol sequence with size equal to the number of evaluations made by the new algorithm is also shown in the same figure. Clearly, Sobol is slightly better than MED generated by the new algorithm, but we should remember that the Sobol sequence can be used only when the distribution function is known, which doesn't happen in a real Bayesian problem.

\section{A COMPUTATIONALLY EXPENSIVE EXAMPLE}
\begin{figure}[h]
\begin{center}
\includegraphics[width = 0.45\textwidth]{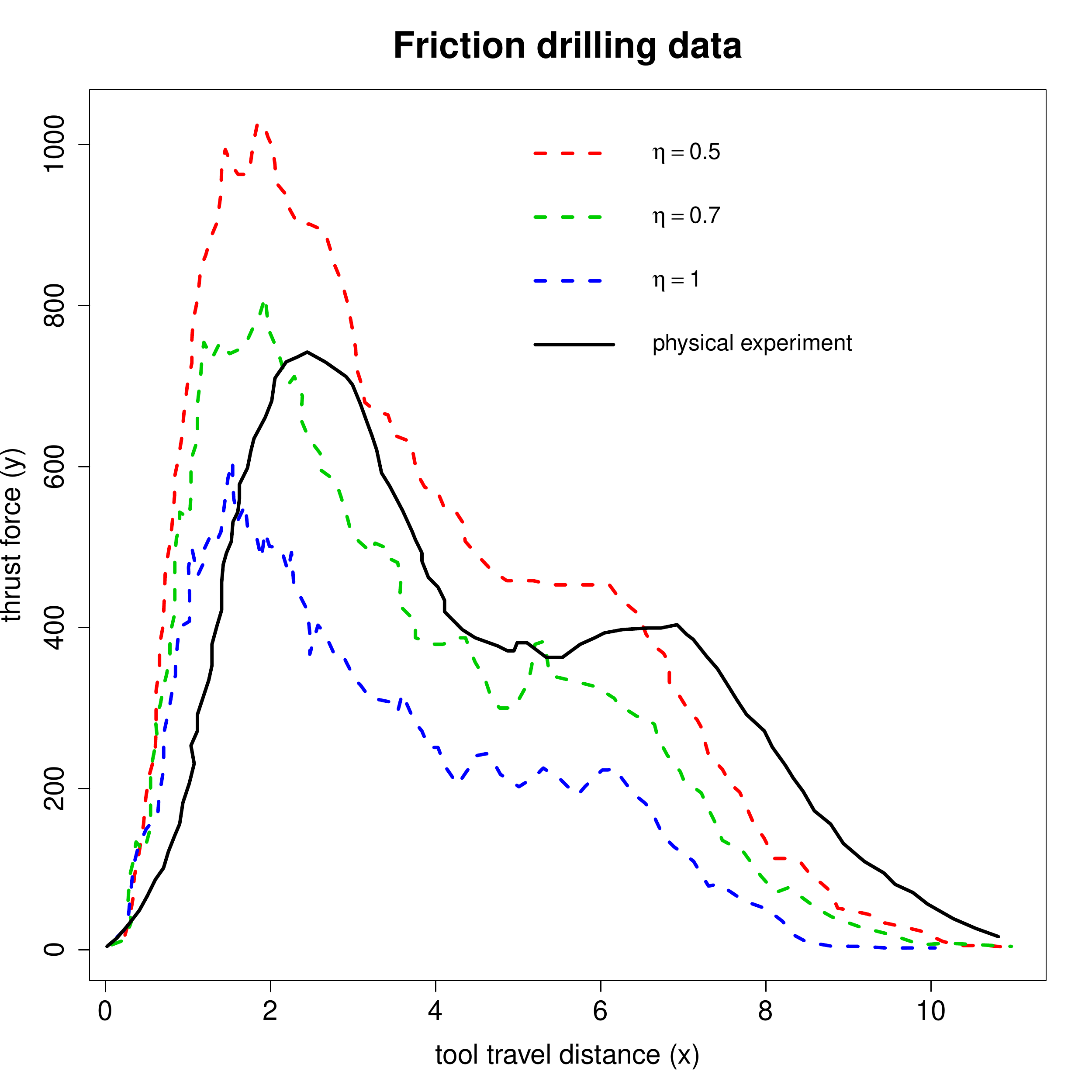}
\caption{FEM outputs at three values of the friction coefficient and actual force data from physical experiment.}
  \label{fig:FEM}
\end{center}
\end{figure}

\citet{Miller2007} developed a thermo-mechanical finite element model (FEM) to simulate a friction drilling process. The model has several outputs, but here we analyze only one output of the model, the thrust force ($y$). The FEM contains an unknown parameter, coefficient of friction ($\eta$), which needs to be specified to get the output. Figure \ref{fig:FEM} shows the thrust force over the tool travel distance ($x$) for three different values of the coefficient of friction. Miller and Shih also performed a physical test to validate the FEM model. The measured thrust force from the physical test is also shown in Figure \ref{fig:FEM}. From this, they concluded that $\eta=0.7$ is the best choice for the friction coefficient. However, they also noticed the discrepancy in the thrust force predictions even with the best choice of $\eta$.

\citet{Kennedy2001} proposed a Bayesian calibration approach to simultaneously estimate $\eta$ and the discrepancy. The discrepancy in their approach is modeled using a Gaussian process (GP). However, a more engineering approach to deal with the discrepancy is to understand the causes of discrepancy and improve the FEM to avoid the discrepancies.
Further investigation of \citet{Miller2007} showed that this discrepancy can be attributed to the deflection of the sheet at the initial contact with the tool. Thus the actual tool travel is  less than the tool travel used in the FEM. It is not easy to implement this correction in the FEM code because the elastic deflection and thrust force need to be solved iteratively which can substantially increase the computational time. \citet{Joseph2015c} proposed to use simple statistical adjustments in such situations. Their approach can be described as follows.

Let $y=g(x;\eta)$ be the FEM. Then, we can take $y=g(\gamma x;\eta)$, where $\gamma \in [0,1]$ is an adjustment parameter, which accounts for the workpiece deflection at the initial contact. If we feel that the deflection can change during tool travel, we can introduce one more adjustment parameter to obtain $y=f(\gamma_1 x^{\gamma_2};\eta)$, where $\gamma_2>1$ indicates that the deflection increases with tool travel. Thus the model calibration problem reduces to the following nonlinear regression problem:
\[y_i=g(\gamma_1 x_i^{\gamma_2};\eta)+\epsilon_i,\]
where $\epsilon_i\sim^{iid}N(0,\sigma^2)$. However, different from the usual nonlinear regression problems, $g(\cdot,\cdot)$ is a computationally expensive model. Therefore, first we approximate the FEM using a GP model:
\[\hat{g}(x;\eta)=\exp\{\hat{\mu}+\bm r(\bm x;\bm \eta)'\bm R^{-1}(\log \bm y^{FEM}-\hat{\mu}\bm 1)\},\]
where $\hat{\mu}=\bm 1'\bm R^{-1}\log \bm y^{FEM}/\bm 1'\bm R^{-1}\bm 1$ and $\bm r(\bm x;\bm \eta)$ and $\bm R$ are the correlation vector and  correlation matrix computed using the Gaussian correlation function $R(\bm h)=\exp(-\theta_x h_x^2-\theta_{\eta} h_{\eta}^2)$. A log-transformation was used to ensure that the predictions are non-negative. The details of GP models can be found in many references, for example, the book by \citet{Santner2003}. We have used the R package \emph{GPfit} \citep{GPfit2015} for fitting the model.

Now to estimate the friction coefficient and the two adjustment parameters, we used the following Bayesian model:
\begin{eqnarray*}
y_i&\sim^{iid} & N\left(\hat{g}(\gamma_1 x_i^{\gamma_2};\eta),\sigma^2\right),\;i=1,\ldots,N,\\
\eta &\sim & p(\eta;.5,1,10,10),\\
\gamma_1&\sim & p(\gamma_1;.5,1,10,100),\\
\gamma_1&\sim & p(\gamma_2;.75,1.25,10,10),\\
p(\sigma^2)&\propto & 1/\sigma^2,
\end{eqnarray*}
where $N=96$ and $p(x;a,b,\lambda_a,\lambda_b)=\exp\{\lambda_a (x-a)\}I(x<a)+I(a\le x\le b)+\exp\{-\lambda_b (x-b)\}I(x>b)$ is the prior distribution. This prior distribution places uniform prior in the interval $[a,b]$ and exponential distributions outside $[a,b]$. Thus, $p(\eta;.5,1,10,10)$ ensures that the estimates of $\eta$ will most likely be in the experimental range of $[.5,1]$, but allows for slight extrapolation outside $[.5,1]$. A larger value of $\lambda_b$ is used for $\gamma_1$ because it is unlikely to be above 1. For simplicity, we have ignored the uncertainties in the estimation of $g(\cdot;\cdot)$.

There are four unknown parameters in the Bayesian model: $(\eta,\gamma_1,\gamma_2,\sigma^2)$ and their posterior distribution is
\[p(\eta,\gamma_1,\gamma_2,\sigma^2|\bm y)\propto \frac{1}{\sigma^N}\exp\left[-\frac{1}{2\sigma^2}\sum_{i=1}^N \{y_i-\hat{g}(\gamma_1 x_i^{\gamma_2};\eta)\}^2\right]p(\eta)p(\gamma_1)p(\gamma_2)p(\sigma^2).\]
We can easily integrate out $\sigma^2$ to obtain (omitting the normalizing constant)
\[\log p(\eta,\gamma_1,\gamma_2|\bm y)=-\frac{N}{2}\sum_{i=1}^n \{y_i-\hat{g}(\gamma_1 x_i^{\gamma_2};\eta)\}^2+\log p(\eta)+\log p(\gamma_1)+\log p(\gamma_2).\]
There were 332 observations in the FEM data, so the prediction using the GP model still takes time. One prediction using \emph{GPfit} took about 0.16 seconds in a 3.2GHz laptop. 96 such evaluations are needed to compute  $\log p(\eta,\gamma_1,\gamma_2|\bm y)$. Thus, one evaluation of the unnormalized posterior takes about 15.4 seconds.

\begin{figure}
\begin{center}
\begin{tabular}{ccc}
\includegraphics[width = 0.3\textwidth]{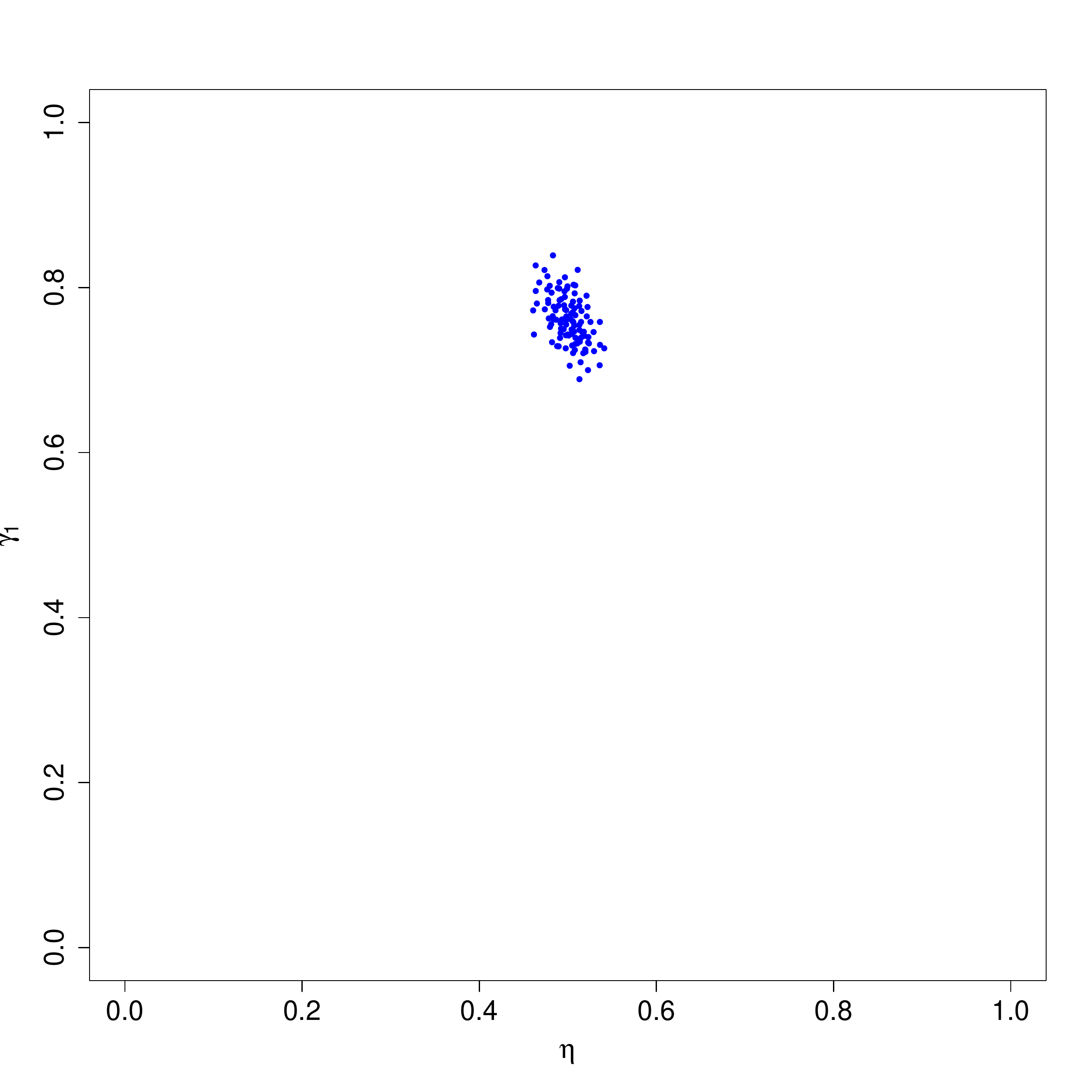} &
\includegraphics[width = 0.3\textwidth]{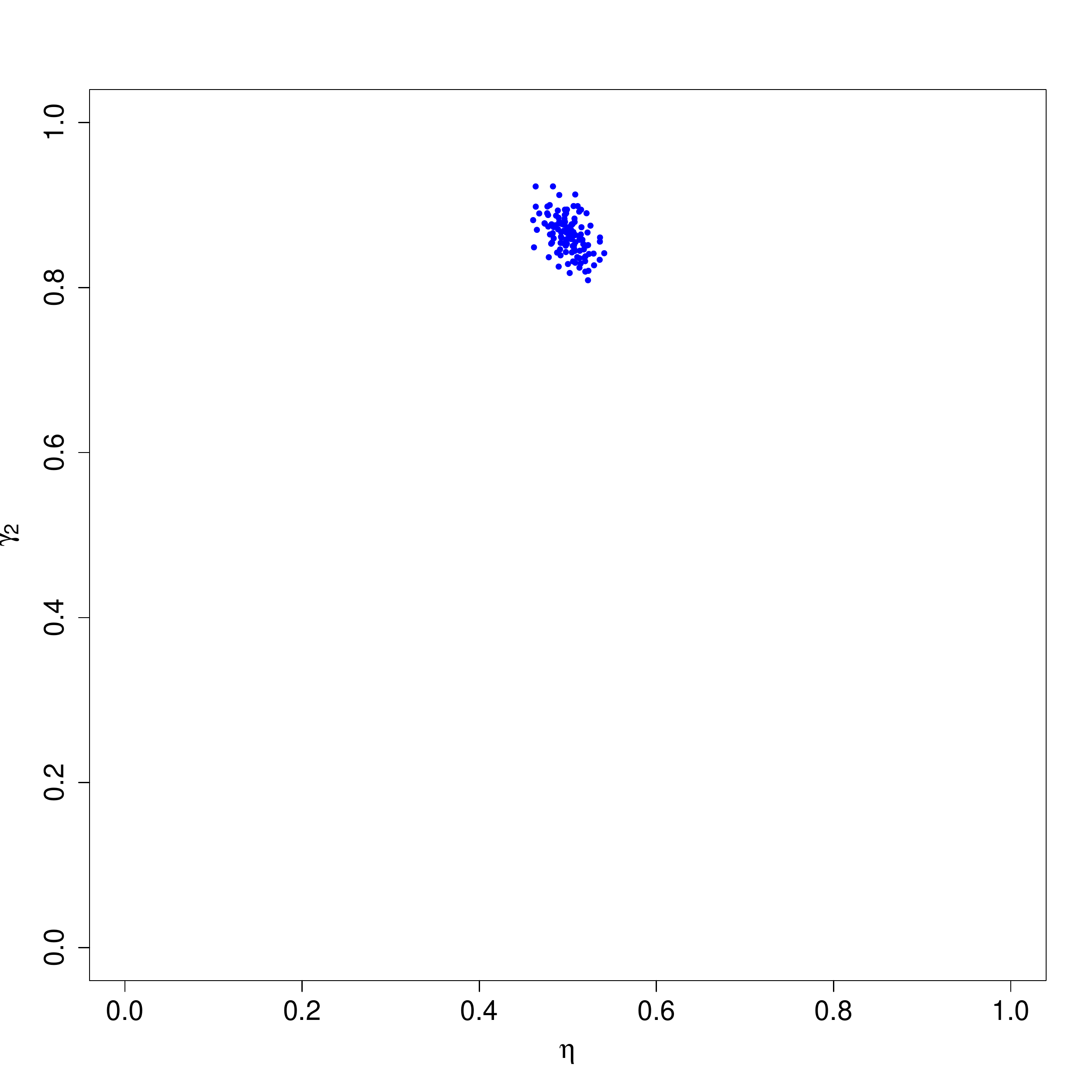} &
\includegraphics[width = 0.3\textwidth]{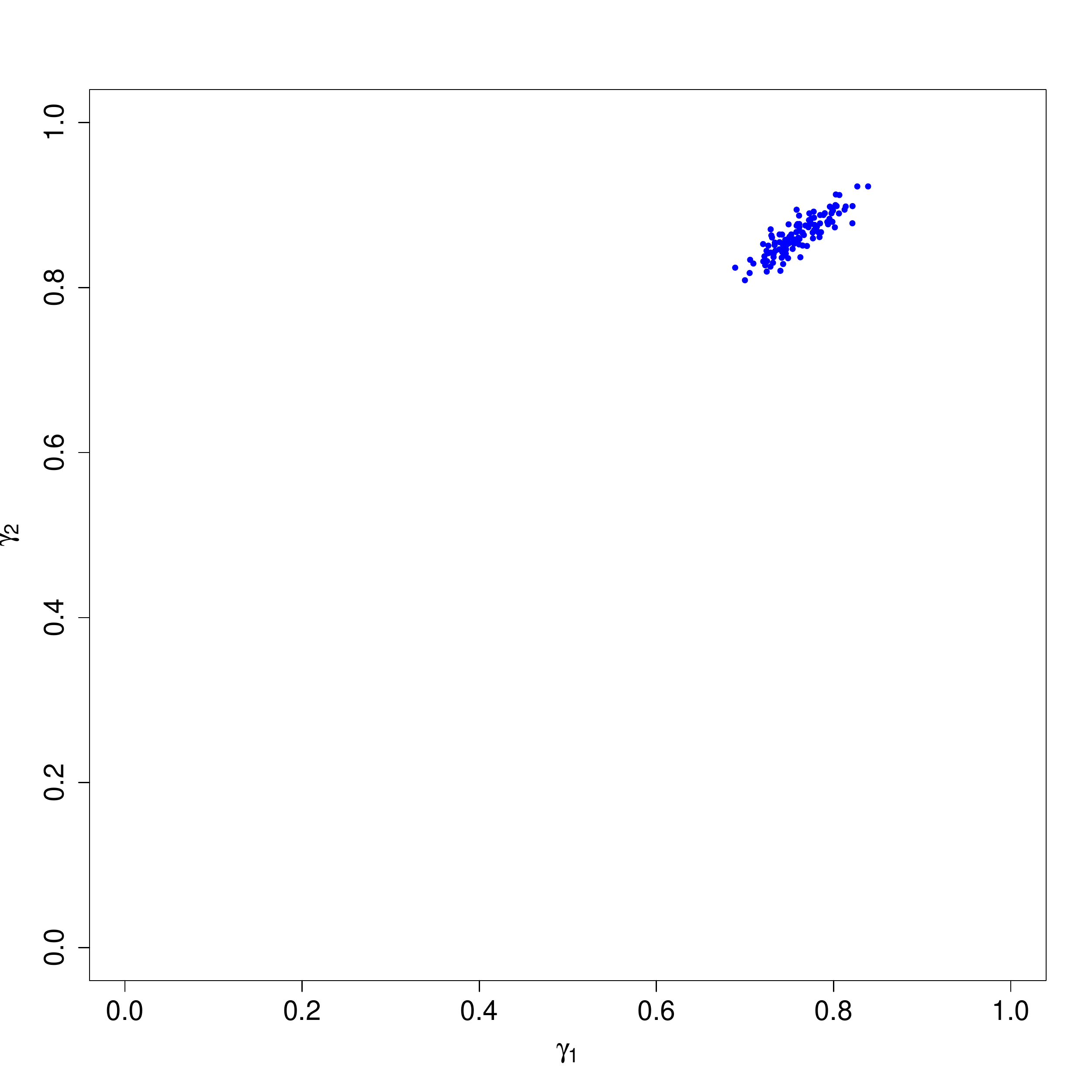}
\end{tabular}
\caption{Two-dimensional plots of the MED points in the friction drilling example.}
\label{fig:FD}
\end{center}
\end{figure}

We scale the parameters in $[0,1]^3$ and apply the algorithm proposed in Section 3 to generate an MED of size 113 with $K=7$. The two-dimensional projections of the MED points are shown in Figure \ref{fig:FD}. We can see that the posterior distribution occupy very little space in the 3-dimensional hypercube, so a QMC sample would have been quite wasteful. On the other hand, it took only about 3.4 hours to generate the MED points. A typical application of MCMC for this problem would require about 10,000 samples, which would have taken about 43 hours to complete. Clearly, MED is advantageous in this particular example.

\begin{figure}[h]
\begin{center}
\includegraphics[width = 0.45\textwidth]{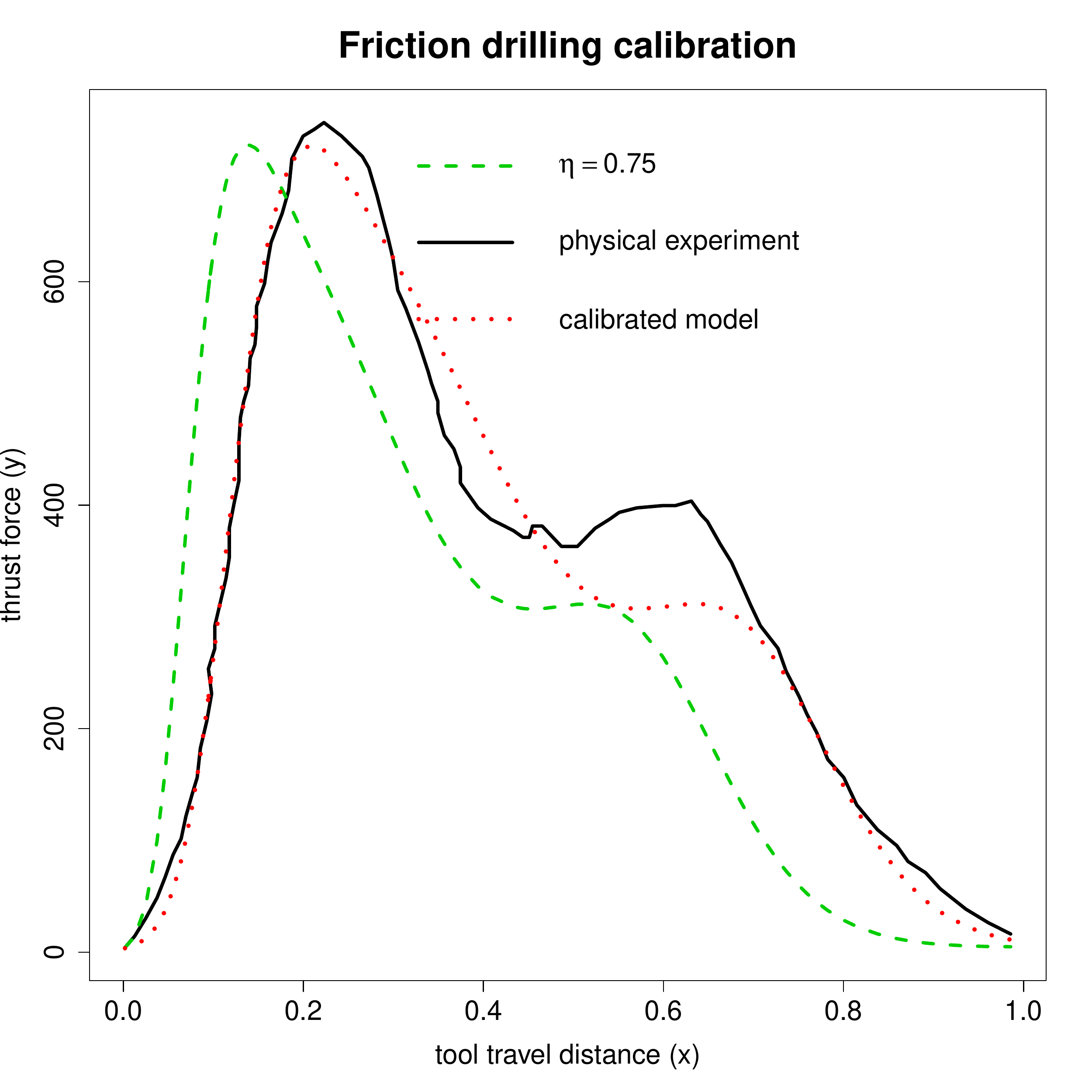}
\caption{Predicted FEM output at $\eta=0.75$, measured data from the physical experiment, and the calibrated model at the posterior means of the parameters.}
  \label{fig:calibration}
\end{center}
\end{figure}

The posterior means of the parameters were $(0.5,0.761,0.863)$ in the $0-1$ scale. In the original scale, they are $\hat{\eta}=0.75$, $\hat{\gamma}_1=0.88$, and $\hat{\gamma}_2=1.18$. The computer model prediction at $\hat{\eta}=0.75$, the measured data, and the prediction from the calibrated model are shown in Figure \ref{fig:calibration}. We can see that the calibration has helped in bringing the computer model prediction closer to the actual measured data. There is still some unaccounted discrepancy. If this is considered important, we may add a GP term to capture the remaining discrepancy.

\section{CONCLUSIONS}
This article proposed an efficient algorithm to generate a minimum energy design. The main novelty is in generating the design with as few evaluations as possible of the probability density, which can be advantageous when dealing with computationally expensive posterior distributions in Bayesian applications. For a 30-dimensional distribution, the new algorithm required 1500 times fewer evaluations compared to the global optimization-based algorithm used in \citet{Joseph2015a}. This substantial reduction in the number of evaluations makes the new algorithm useful in real applications and competitive to the existing MCMC and QMC methods. The algorithm can work as a black-box and the user need to specify only two inputs: the log-unnormalized posterior and a hypercube where the density is expected to lie. Thus it can be used easily by non-statisticians and applied to a wide variety of Bayesian problems. We have made sensible choices to various parameters involved in the algorithm, which seem to work well in the problems we have tested so far. Clearly, much more needs to be done to understand the convergence of the algorithm and optimal choices of the parameters, but we leave this as a topic for future research.

It is important to understand when to use a deterministic sampling method such as MED as opposed to MCMC in a real Bayesian application. It is our belief that practitioners will continue to use MCMC because of its flexibility and ease of implementation, as long as the likelihood and prior are cheap to evaluate. But as the cost of evaluations of the unnormalized posterior increases, the deterministic sampling methods become more relevant. On the other hand, if the cost of evaluation is very high, then even the deterministic sampling can become unaffordable and one may need to rely on more efficient function approximation techniques. The decision to transition from MCMC to MED to function approximation depends on the problem in hand, the available computing resources, the time constraints, and the accuracy needed. We admit that the accuracy of the deterministic samples or function approximation can be questionable for complex distributions and in high dimensions. The accuracy produced by them may be enough for the problem in hand, particulary at the initial stages of model building. But if high accuracy is warranted, then one may need to follow-up such a method with MCMC (see Figure \ref{fig:bananax1x2}). On the other hand, the deterministic samples can be used to make the MCMC methods more efficient or can act as an experimental design for doing function approximation. Thus, in summary, MED is not meant to replace MCMC or other advanced function approximation techniques, but it can play a major role in solving a Bayesian problem when the computations are expensive.

\begin{center}
{\Large\bf SUPPLEMENTARY MATERIALS}
\end{center}

\noindent The proof of Theorem 1 is given in this file.\\

\begin{center}
    {\Large\bf ACKNOWLEDGMENTS}
\end{center}
This research is supported by a U.S. National Science Foundation grant DMS-1712642.

\bibliographystyle{ECA_jasa}
\bibliography{Ref_all}

\end{document}